\documentclass[aps,pra,superscriptaddress,twocolumn]{revtex4}
\usepackage{graphicx,amsmath,amssymb,amsfonts,latexsym,color,dcolumn,bm}

\newcommand{\beq}{\begin{equation}}
\newcommand{\eeq}{\end{equation}}
\newcommand{\bea}{\begin{eqnarray}}
\newcommand{\eea}{\end{eqnarray}}
\providecommand{\abs}[1]{\left\lvert#1\right\rvert}

\providecommand{\bra}[1]{\langle #1 \rvert}
\providecommand{\ket}[1]{\lvert #1 \rangle}

\usepackage{amsfonts,amssymb}
\usepackage{dsfont}
\usepackage{physics}
\usepackage{tocvsec2}

\usepackage{appendix}

\usepackage{tikz}

\usepackage{amsmath}
\usepackage{bbold}
\usepackage{hyperref}
\hypersetup{
     colorlinks=true,      
    linkcolor=blue,        
    citecolor=blue,        
    filecolor=magenta, 
    urlcolor=black          
}

\usepackage{comment}

\begin{document}

\title{Wigner distribution on a double cylinder phase space for studying quantum error correction protocol}
\author{N. Fabre\footnote{nicolas.fabre@univ-paris-diderot.fr}}
\affiliation{Laboratoire Mat\'eriaux et Ph\'enom\`enes Quantiques, Sorbonne Paris Cit\'e, Universit\'e de Paris, CNRS UMR 7162, 75013 Paris, France}
\author{A. Keller}
\affiliation{Laboratoire Mat\'eriaux et Ph\'enom\`enes Quantiques, Sorbonne Paris Cit\'e, Universit\'e de Paris, Univ. Paris Saclay, CNRS UMR 7162, 75013 Paris, France}
\author{P. Milman}
\affiliation{Laboratoire Mat\'eriaux et Ph\'enom\`enes Quantiques, Sorbonne Paris Cit\'e, Universit\'e de Paris, CNRS UMR 7162, 75013 Paris, France}

\date{\today}
\begin{abstract}
 We introduce a quasi-probability phase space distribution with two pairs of azimuthal-angular coordinates. This representation is well adapted to describe quantum systems with discrete symmetry. Quantum error correction of states encoded in continuous variables using translationally invariant states is studied as an example of application. We also propose an experimental scheme for measuring such new distribution.
 \end{abstract}
\vskip2pc

\maketitle


\section{Introduction}

Quantum information can be encoded in discrete variables (DV) - in a two-dimensional (or higher) Hilbert space - or in continuous variables (CV) - in an infinite dimensional Hilbert space - of physical systems. The CV can be for instance the two canonical conjugate quadratures of the electromagnetic field \cite{CVQC, CV2}, phonons or the continuous degrees of freedom at the single boson level, such as time-frequency \cite{fabregkp} or the transversal ones \cite{spatialcv}. Quantum computation involves manipulating states encoded in CV or in DV using an universal set of unitary gates. To achieve quantum advantage with either type of encoding \cite{SpeedupCVQC}, such universal set must possess a non-Clifford gate, which cannot be efficiently simulated using classical ressources. Quantum computation using DV or CV can achieve exponential speedup for specific tasks, as for integer factoring \cite{shorfactor} or quantum simulation \cite{feymansimulation,Lloydsimulation}. The possibility of scalability has been demonstrated by the experimental realization of multimode CV cluster states \cite{Cluster1,CVcluster}. Nevertheless, quantum error correction (QEC) and fault tolerance require finite dimensional Hilbert spaces. 
 Bosonic codes constitute a solution for defining QEC using CV encoding: they are a class of quantum codes where logical qubits are encoded by defining a protected finite dimensional subspace within an infinite-dimensional Hilbert space. One example of such codes is the so-called cat-code, which was the first one to be introduced \cite{Bosoniccat} and possesses parity symmetry.  The logical zero and one qubits correspond to the two possible parity eigenstates. Photon losses causes errors that are equivalent to a bit flip, which can be corrected with a quantum error correction circuit \cite{correctcat1}.
An alternative solution for QEC is to encode information redundantly, using quantum states with translational symmetry in a rectangular phase space.  This was introduced by Gottesman, Kitaev and Preskill (GKP) \cite{GKP}. One can also use rotational symmetry as introduced in \cite{rotational,rotational3}. The GKP states, also called grid states, are CV states composed of a finite number of position localized states, which can be represented by peaks in the probability density distribution. The widths of both the peaks and of the envelope determining the extent of the superposition (number of peaks) depend on the number of photons of the state {\it{i.e}} on the squeezing. Narrow peaks and a wide envelope mean that the state has a high level of squeezing. Initial squeezing of 20.5 dB in a cluster composed of GKP states allows to a fault-tolerant measurement based quantum computation as shown in \cite{Menicu1}. These type of bosonic codes are designed to be robust against errors which are translational (or rotational) displacement in phase space. As demonstrated in \cite{Albert}, GKP codes are also robust to protect against pure-loss channel and over perform the cat codes for a high number of photons, even though the cat codes are designed to protect against such errors. In addition, the deep interest of such states is motivated  from its recent experimental realization in different platforms, such as superconducting cavities \cite{supercondgkp}, the time-frequency variables at the single photon level in an integrated photonic circuit \cite{fabregkp} and trapped ions \cite{TrappedionsGKP}.

Phase space quasi-probability distributions are a valuable tool for understanding and visualizing quantum states. Stratonovich was the first to use an axiomatic approach to build a distribution in phase space \cite{stratotrue}. Such distribution, like bosonic codes, are constructed on symmetry considerations. Therefore, symmetries can be particularly adapted to describe each code. For instance, when the dynamical group is a $\mathbb{SU}(2)$ symmetry \cite{spindistri,wignersu2}, the phase space is a $2$-sphere whereas for the Heisenberg-Weyl  and the cylindrical group, the phase space is rectangular and cylindrical respectively.

Defining two lattices along the position and momentum variables with periodicity $l$ and $2\pi/l$,  we can define modular variables (MV). MV are a specific class of observables that are periodic and have been introduced in \cite{Aharo} to explain non-local properties of quantum states. They are now used in quantum information protocol to identify discrete structure in continuous variable states \cite{Keterrermodular,Modularexp,carvalho}.  MV are particularly adapted in situations where the considered quantum state has a periodic structure. We also note the recent  interest of this formalism among the high energy community for understanding quantum gravity or metastring theory\cite{metastring,pathintegral,modularhigh}.

We propose in this paper to study systems which have translational symmetry. For such, we define a four dimensional phase space and  define a quasi-probability  distribution, obeying Stratonovich-Weyl  \cite{strato} rules, with two pairs of azimuthal-angular variables. The associated angular variables correspond to the MV. It leads to the construction of a double cylinder phase space. We then generalize  the results obtained on the rotational Wigner distribution on a single cylinder phase space, as  introduced in \cite{rigas,cylinder,cylinder1,bizarro}, or for the phase-number Wigner distribution one \cite{Wignernumber1,Wignernumber2}.  The representation of a quantum state  in a double cylinder phase space is  totally equivalent to the rectangular phase space. But it turns out to be the most adapted one for states  exhibiting translational symmetry. We show in two specific examples that codes which exhibit the same symmetry as the one underlying the construction of a phase space have a similar Wigner distribution in its corresponding phase space. In particular, Schrödinger cat codes which have parity symmetry are represented in the rectangular phase space by two peaks and an oscillation pattern perpendicular to the peaks. Remarkably GKP states, which have translational symmetry have exactly the same shape in a double cylindrical phase space, which we introduce in this paper.  We further propose to study a quantum error correction protocol relying on the use of a position-momentum GKP state as ancilla  and visualize the correction into this phase space.

The paper is organized as follows. In Sec. \ref{sectiontwo}, we recall the modular variables formalism and their properties. In Sec. \ref{sectionthree}, the properties of geometrical modular quantization are introduced emphasizing the need of a double cylinder phase space. 
In Sec. \ref{sectionfour}, we use this new representation to study the GKP states, which are translationally invariant states in the usual rectangular phase space. In Sec. \ref{sectionfive}, we represent the GKP,  the coherent and cat state in a double cylinder phase space. We then recall the principles of quantum error correction in Sec. \ref{sectionsix} of GKP and Gaussian state using as ancilla a GKP state. We give a new figure of merit for the probability of having an error in this encoding. Finally in Sec. \ref{sectionseven}, we propose an experimental scheme to measure such distribution.

\section{Modular variables formalism}\label{sectiontwo}

In this section, we describe an alternative representation  of  quantum states using modular variables, which is well adapted for physical systems with translational symmetry. For such, we will introduce the Zak's transform, an operation that takes as input a function of one variable (which belongs to $\mathds{R}$) and produces as output a function of two variables (which belong to the torus $\mathds{S}^{1}\cross \mathds{S}^{1 *}$). The modular basis and its canonically conjugate one are defined. In the following, we set $\hbar=1$ but it will be reintroduced at the end of the calculation.

\subsection{Modular basis and the Zak's transform}
\begin{figure}[h]
\begin{center}
 \includegraphics[scale=0.17]{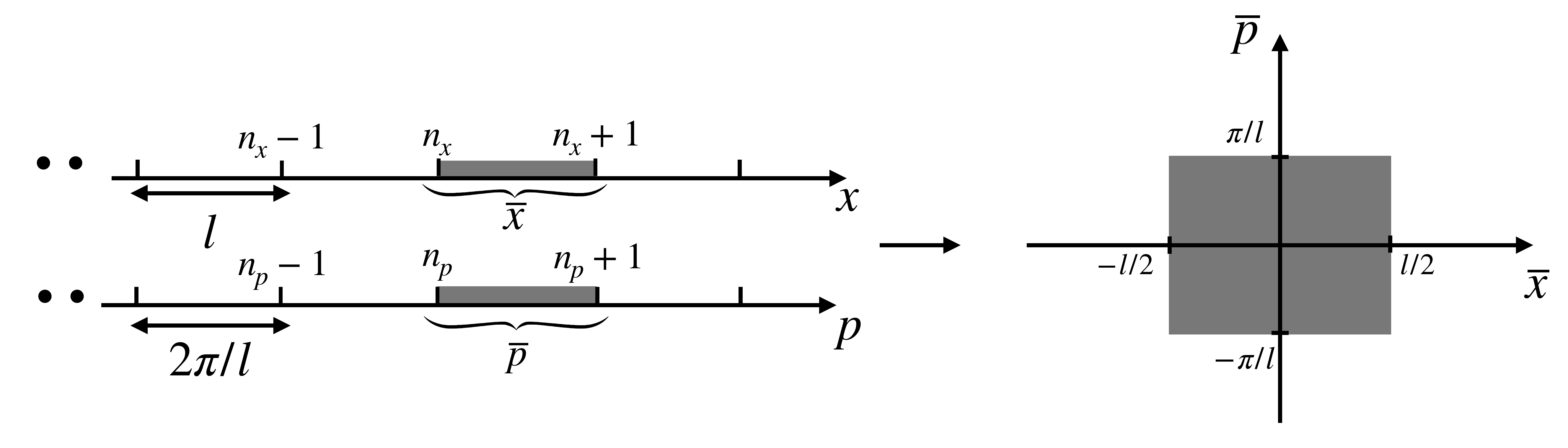}
 \caption{\label{latticevariablemodular} Schematic representation of the position and momentum lattices, with period $x_{0}=l$ and $p_{0}=2\pi/l$. The integers $n_{x}$, $n_{p}$ label the cells along the $x$,$p$ axis and the modular variables $\overline{x},\overline{p}$ correspond to the position and momentum values inside one cell.}
 \end{center}
\end{figure}

As shown in \cite{Aharonov,Keterrermodular}, the pairs of canonically conjugate variable $\hat{x}$ and $\hat{p}$, related by the commutation relation $[\hat{x},\hat{p}]=i$ can be decomposed into :
\begin{equation}
\hat{x}=\hat{N}_{x}x_{0}+\hat{\overline{x}}\ ,\ \hat{p}=\hat{N}_{p}p_{0}+\hat{\overline{p}},
\end{equation}
where $\hat{N}_{x}$ and $\hat{N}_{p}$ are operators with integers eigenvalues $n_{x}$ and $n_{p}$ which define a lattice along the $x$ and $p$-axis, with period $x_{0}$ and $p_{0}$ respectively (see Fig.~\ref{latticevariablemodular}). $\overline{\hat{x}}$ and $\overline{\hat{p}}$ are bounded operators which spectra lie on the intervals $\frac{1}{2}[-x_{0},x_{0}[$,  $\frac{1}{2}[-p_{0},p_{0}[$ respectively. In order to quantize the phase space, as we will see, it is not possible to have operators which are discontinuous as the bounded operator $\overline{\hat{x}}$ and $\overline{\hat{p}}$. Instead, we can choose smooth versions of these operators, such as their cos, sin or the exponential versions \cite{smoothoperator}. The exponential form corresponds to displacement operators defined by $\hat{\cal{D}}(x,p)=e^{i(\hat{x}p-\hat{p}x)}$. 
It was shown in \cite{Busch1,Busch2} that if $x_{0}p_{0}=2\pi$, the commutator of the two displacements operators vanishes: 
\begin{equation}\label{commutatordis}
[\hat{\cal{D}}(2\pi,0),\hat{\cal{D}}(0,2\pi)]=0,
\end{equation}
 Hence, the quantum algebra of modular variable differs from the classical algebra one, since the corresponding Poisson brackets of Eq.~(\ref{commutatordis}) is not zero \cite{metastring}.
In the following, we choose the values $x_{0}=l$ and $p_{0}=\frac{2\pi}{l}$. When the previous condition (see Eq.~(\ref{commutatordis})) is fulfilled, we can define the modular basis $\ket{\overline{x},\overline{p}}$, which are the common eigenstates of the displacement operators:
\begin{multline}\label{boundedoperator}
\exp(i\hat{\overline{x}}\mu)\ket{\overline{x},\overline{p}}=\exp(i\overline{x}\mu)\ket{\overline{x},\overline{p}},\\
\exp(i\hat{\overline{p}}\alpha)\ket{\overline{x},\overline{p}}=\exp(i\overline{p}\alpha)\ket{\overline{x},\overline{p}},\\
\end{multline}
for  $\mu\in\frac{1}{2}[-\frac{\pi}{l},\frac{\pi}{l}[$ and $\alpha\in \frac{1}{2}[-l/2,l/2[ $. The MV basis is orthogonal since we have: $\bra{\overline{x}',\overline{p}'}\ket{\overline{x},\overline{p}}=\delta_{l}(\overline{x}-\overline{x}')\delta_{\frac{2\pi}{l}}(\overline{p}-\overline{p}')$ (where $\delta_{l}$ is the Dirac comb of period $l$) and also satisfy the completeness relation: $\iint \text{d}\overline{x} \text{d}\overline{p} \ket{\overline{x},\overline{p}}\bra{\overline{x},\overline{p}} =\mathds{I}$. 
The integer operators do not commute and verify the relation:
\begin{equation}\label{commutationN}
[\hat{N}_{x},\hat{N}_{p}]=\frac{i}{2\pi}\mathds{I}-\frac{1}{l}[\hat{\overline{x}},\hat{N}_{p}]-\frac{l}{2\pi}[\hat{N}_{x},\hat{\overline{p}}],
\end{equation}
which is a consequence of the commutation relation $[\hat{x},\hat{p}]=i$ \cite{slitsmodular,Keterrermodular}. From this equation, we conclude that we cannot build a basis with the integers eigenvalues of the operators $\hat{N}_{x},\hat{N}_{p}$. The two other commutators which appear in Eq.~(\ref{commutationN}) are different from zero and  are calculated in Appendix \ref{appendixcommu}. They  prevent the construction of a basis built with the common eigenstates of $\hat{N}_{x},\hat{N}_{p}$. Nevertheless, if we do not impose that the period of the $x$-lattice and the period of the $p$-lattice are related as before (periodic boundary condition), it would be possible to build such basis.\\
The Zak's transform  (also called Weyl-Brezin-Zak transform \cite{cylinder}) is a mapping from $L^{2}(\mathds{R})\rightarrow L^{2}(\mathds{S}^{1}\cross \mathds{S}^{1 *})$ (where  $\mathds{S}^{1 *}$ is the dual of the circle $\mathds{S}^{1}$) which permits the representation of a position or momentum function by a two-dimensional function \cite{slitsmodular, Keterrermodular}. For a position state, it can be written in the following form:
\begin{equation}\label{Zakt}
\ket{x}=\ket{\overline{x}+ml}=\int_{-\frac{\pi}{l}}^{\frac{\pi}{l}} \text{d}\overline{p}e^{-im\overline{p}l} \ket{\overline{x},\overline{p}},
\end{equation}
and for the inverse of the Zak's transform,
\begin{equation}
\ket{\overline{x},\overline{p}}=\sum_{m\in\mathds{Z}}e^{im\overline{p}l}\ket{\overline{x}+ml}.
\end{equation}
 In the $p$-representation, the Zak's transform can be written as:
\begin{equation}
\ket{p}=| \overline{p}+\frac{2\pi }{l}n \rangle=e^{-i\overline{x}\cdot\overline{p}}\int_{-l/2}^{l/2} \text{d}\overline{x} e^{2i\pi n \overline{x}/l} \ket{\overline{x},\overline{p}}, 
\end{equation}
The modular basis being orthogonal, we can decompose a wave function $\ket{\psi}=\int \text{d}x \psi(x) \ket{x}$ as:
\begin{align}\label{modularwave}
\ket{\psi}=\int_{-\frac{\pi}{l}}^{\frac{\pi}{l}}\int_{-l/2}^{l/2} \psi(\overline{x},\overline{p}) \ket{\overline{x},\overline{p}} \text{d}\overline{x} \text{d}\overline{p}, \\ \psi(\overline{x},\overline{p}) = \sum_{n\in\mathds{Z}} \psi(\overline{x}+nl)e^{-in\overline{p}l}.
\end{align}
$\psi(\overline{x},\overline{p})$ is called the modular wave function and can be represented in a torus $\mathds{S}^{1}\cross \mathds{S}^{1 *}$.
The modular wave function is quasi-periodic in $\overline{x}$ and periodic in $\overline{p}$:
\begin{align}
\psi(\overline{x}+l,\overline{p})=e^{il\overline{p}}\psi(\overline{x},\overline{p}),\\
\psi(\overline{x},\overline{p}+\frac{2\pi}{l})=\psi(\overline{x},\overline{p})
\end{align}
The modular wave function is uniquely defined by its restriction to the unit square (the unfolded torus). The quasi-period is reminiscent of the Aharanov-Bohm potential \cite{Ahabohm}.

\subsection{Fourier transform of the modular variable basis}\label{Modularbasis}

The canonical conjugate basis of the modular basis is defined by the double Fourier series \cite{Zak},$ \{ (\overline{x}, \overline{p})/ \overline{x}\in \mathds{S}^{1},  \overline{p}\in \mathds{S}^{1*} \}\rightarrow \{ (\frac{2\pi}{l} n,ml) / n,m\in \mathds{Z}\} $: 
\begin{equation}\label{modtoint}
\ket{\overline{x},\overline{p}}=\sum_{n,m\in\mathds{Z}^{2}} | n,m \rangle e^{-i(\frac{2\pi}{l}n\overline{x}-ml\overline{p})}.
\end{equation}
is called the integer basis or the discrete Zak's basis. The inverse relation is the Fourier transform of the two bounded integrals:
\begin{equation}\label{inttomod}
\ket{n,m}=\int_{-\frac{l}{2}}^{\frac{l}{2}}\int_{-\frac{\pi}{l}}^{\frac{\pi}{l}} \text{d}\overline{x} \text{d}\overline{p} e^{i(\frac{2\pi n}{l}\overline{x}-ml\overline{p})} \ket{\overline{x},\overline{p}}
\end{equation}
This basis is also orthogonal since $\bra{n',m'}\ket{n,m}=\delta_{n,n'}\delta_{m,m'}$ and satisfy the closure relation $\sum_{n,m\in\mathds{Z}^{2}} \ket{n,m}\bra{n,m}=\mathds{I}$. The scalar product of the two canonically conjugate basis is:
\begin{equation}
\bra{n,m}\ket{\overline{x},\overline{p}}=e^{i(\frac{2\pi n}{l}\overline{x}-ml\overline{p})} ,
\end{equation}
which means that they are mutually unbiased.
We define the two integers operators $\hat{N},\hat{M}$ with discrete spectrum $n,m\in\mathds{Z}^{2}$:
\begin{equation}\label{integeroperator}
\hat{N}\ket{n,m}=n\ket{n,m}\ , \ \hat{M}\ket{n,m}=m\ket{n,m},
\end{equation}
 and they necessarily commute $[\hat{N},\hat{M}]=0$. Thus, we cannot identify these operators to $\hat{N}_{x},\hat{N}_{p}$. Even though we can know with an infinite precision the eigenvalues of the bounded operators, $\hat{\overline{x}}$, $\hat{\overline{p}}$, $\hat{N}$ and $\hat{M}$ this is not the case for the integers eigenvalues of $\hat{N}_{x},\hat{N}_{p}$. They cannot be known simultaneously as a consequence of Heisenberg's inequality  \cite{Zak}.\\

The only nonzero commutator between the four operators Eq.~(\ref{boundedoperator}) and Eq.~(\ref{integeroperator})  are: $[\hat{N},\exp(i\hat{\overline{x}}\mu)]= \mu\exp(i\hat{\overline{x}}\mu)$ and $[\hat{M},\exp(i\hat{\overline{p}}\alpha)]=\alpha\exp(i\hat{\overline{p}}\alpha)$, with similar results in \cite{bizarro,ordering}. Hence $(\overline{x},n)$ and $(\overline{p},m)$ form a couple of independent azimuth-angular variables. We can hence write the wave function in the integer's basis:
\begin{equation}\label{integerwavefunction}
\ket{\psi}=\sum_{n,m\in \mathds{Z}^{2}} \psi_{n,m}\ket{n,m}.
\end{equation}
The relations between the modular wave function and the integer wave function are:
\begin{align}\label{integerwavefunction}
\psi_{n,m}=\int_{-\frac{\pi}{l}}^{\frac{\pi}{l}}\int_{-l/2}^{l/2} \psi(\overline{x},\overline{p})e^{-i(\frac{2\pi n}{l}\overline{x}-m\overline{p}l)}\text{d}\overline{x}\text{d}\overline{p}, \\ \psi(\overline{x},\overline{p})=\sum_{n,m\in\mathds{Z}^{2}}e^{i(\frac{2\pi n}{l}\overline{x}-m\overline{p}l)}\psi_{n,m}.
\end{align}
The integer wave function  defined by Eq.~(\ref{integerwavefunction}) can be seen as a bipartite qudit state. $\psi_{n,m}$ is non-separable if it cannot be written under the form $\psi_{n,m}=f_{n}g_{m}$. Usual tools for understanding and quantifying non-separability of pure bipartite qudit system, as for instance the Schmidt decomposition, can be used in this context \cite{Nielsen}. \\

 The transformation relating the $x$-representation to the integer-representation is:
{\small{\begin{equation}\label{integertox}
\ket{x}=\ket{\overline{x}+ml}=\sum_{n,m'\in\mathds{Z}^{2}}\text{sinc}\left( \frac{(m-m')\pi}{\hbar}\right)e^{-in2\pi\overline{x}/l}\ket{n,m'}
\end{equation}}}
and the inverse relation is:
\begin{equation}\label{relationinteger}
   \ket{n,m}= \int_{-l/2}^{l/2}\sum_{m'\in\mathds{Z}} e^{in2\pi\overline{x}/l}\text{sinc}\left(\frac{(m-m')\pi}{\hbar}\right)\ket{\overline{x}+m'l} \text{d}\overline{x},
\end{equation}
where we used the formula $\int_{-\pi/l}^{\pi/l} e^{-inl\overline{p}}e^{iml\overline{p}} \text{d}\overline{p}=\pi\hbar \text{sinc}\left(\frac{(m-n)\pi}{\hbar}\right)$. We can also find the relation from the $p$-representation to the integer one in an analogous way.
Thanks to these relations, it becomes more clear why it is not possible to identify the integer which labels the position in the Hilbert space $x$ here noted $m'$ in Eq.~(\ref{relationinteger}) and the  integer $m$ which is the eigenvalue of the angular momentum operator $\hat{M}$.\\

\begin{figure}[h]
\begin{center}
 \includegraphics[scale=0.45]{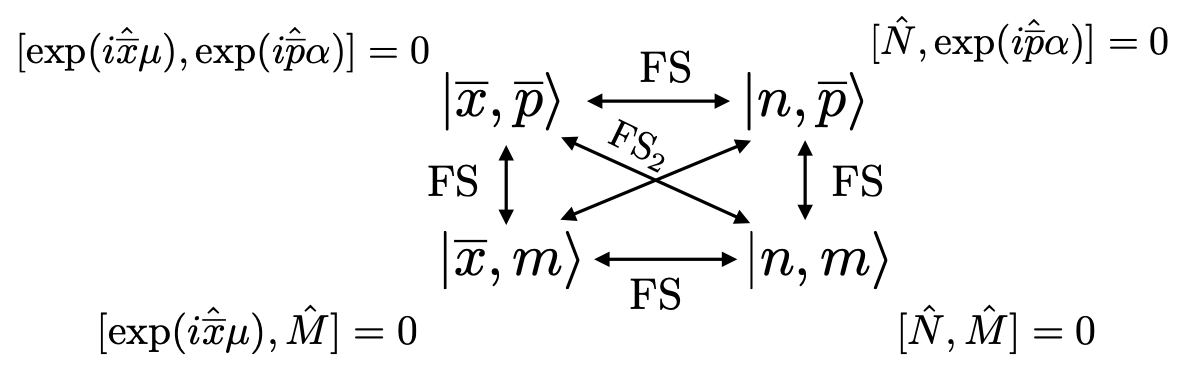}
 \caption{\label{modular}Relation between the different choices of basis. FS stands for Fourier series and $\text{FS}_{2}$ for double Fourier series.}
 \end{center}
\end{figure}

We can also define two other  hybrid basis with two variables, where one of them is bounded and the other is an integer, as $\ket{\overline{x},m}$ or $\ket{n,\overline{p}}$. They constitute an orthogonal basis such that $\bra{\overline{x},n}\ket{\overline{x}',n'}=\delta(\overline{x}-\overline{x}')\delta_{n,n'}$ and they are also complete. The two hybrid basis are mutually unbiased since:
\begin{equation}
\bra{\overline{x},m}\ket{n,\overline{p}}=e^{i(\frac{2\pi n}{l}\overline{x}-ml\overline{p})}.
\end{equation}
The wave function can be expressed in the basis $\ket{\overline{x},m}$, for instance, as:
\begin{equation}
\ket{\psi}=\sum_{m\in\mathds{Z}}\int_{-l/2}^{l/2} \text{d}\overline{x} \psi(\overline{x},ml) \ket{\overline{x},ml},
\end{equation}
where the amplitude  $\psi(\overline{x},\overline{p})$  is given by
\begin{equation}\label{marginaleq}
 \psi(\overline{x},ml)=\int_{-\pi/l}^{\pi/l}  \psi(\overline{x},\overline{p}) e^{im\overline{p}l} \text{d}\overline{p}
\end{equation}
The other relations between the various wave function can be deduced straightforwardly in an analogous fashion.
The relations between the different choices of basis are presented in Fig.~\ref{modular} together with the different commutation relation between the associated operators.

\section{Quantization in a double cylinder phase plane}\label{sectionthree}
In this section, after introducing the displacement and point operator \cite{paz,wooters} on a double cylinder phase space, we build the  associated Wigner distribution.

\subsection{Displacement operator and phase point operator on a double cylinder}

\subsubsection{Modular Displacement operator}
The modular variables $\overline{x}, \overline{p}$ are angle coordinates and are canonically conjugated to the integers variable $n,m$ which are the angular momentum variable.  We define the modular displacement operator as:
\begin{equation}
\hat{D}(n,\overline{x},m,\overline{p})=e^{i(\overline{x}\frac{2\pi\hat{N}}{l}+\hat{\overline{x}}\frac{2\pi n}{l}-\overline{p}l\hat{M}+\hat{\overline{p}}ml)}.
\end{equation}
Owing to the commutation relations between the different operators (see Fig.~\ref{modular}), the pairs $(n, \overline{x})$ and $(m,\overline{p})$ are dynamical variables of two independent degrees of freedom, as pointed out in \cite{Zak}. The modular displacement operator can hence be written as the product of unitary operators $\hat{D}(n,\overline{x},m,\overline{p})=\hat{D}(n,\overline{x})\hat{D}(m,\overline{p})$, with for instance $\hat{D}(n,\overline{x})=e^{i(\overline{x}\frac{2\pi\hat{N}}{l}+\hat{\overline{x}}\frac{2\pi n}{l})}$.

The modular displacement operator can be rewritten in an integral or sum representation. First, in the integer basis,
\begin{multline}\label{displacementinteger}
\hat{D}(n,\overline{x},m,\overline{p})=\sum_{r,s\in\mathds{Z}^{2}}e^{i\overline{x}(r+n/2)\frac{2\pi}{l}}e^{-i\overline{p}(s+m/2)l}\\ \cross \ket{r+n,m+s}\bra{r,s}.
\end{multline}
And then in the modular basis, 
\begin{multline}
\hat{D}(n,\overline{x})\hat{D}(m,\overline{p})=e^{-in\frac{2\pi}{l}\overline{x}/2}e^{iml\overline{p}l/2} \int_{-l/2}^{l/2}\text{d}\overline{x_{1}}  \int _{-\pi/l}^{\pi/l} \text{d}\overline{p}_{1}\\ \cross e^{i\frac{2\pi}{l}\overline{x_{1}}n} e^{-i\overline{p}_{1}l} \ket{\overline{\overline{x_{1}}+\overline{x}},\overline{{\overline{p}_{1}+\overline{p}}}}\bra{\overline{x_{1}},\overline{p}_{1}}.
\end{multline}
The product of two modular displacement operators is equivalent to a modular displacement operator where the variables are summed up to a phase factor, also called cocycle:
\begin{multline}\label{composition}
\hat{D}(n,\overline{x},m,\overline{p})\hat{D}(n',\overline{x'},m',\overline{p'})e^{-i(\overline{x}n'/2-\overline{x'}n/2}e^{-i(\frac{\overline{p}m'}{2}-\overline{p'}m/2)}\\=e^{\pm i H(\abs{\overline{x}-l\pm \overline{x'}})\frac{(n+n')}{2}}e^{\mp i H(\abs{\overline{p}-\frac{\pi}{l}\pm \overline{p'}})\frac{(m+m')}{2}}\\
\cross\hat{D}(n+n',\overline{\overline{x}+\overline{x'}},m+m',\overline{\overline{p}+\overline{p'}})
\end{multline}
where we used the relation $\overline{\overline{x}+\overline{x'}}=\overline{x}+\overline{x'} \mp l H(\abs{\overline{x}-l/2\pm \overline{x'}})$ for $\overline{x}\gtrless0$ \cite{Mukunda} and  $H(x)$ is the Heaviside function which is zero for negative values of $x$ and equals to one if not. The modular displacement operator forms an orthogonal basis:
\begin{multline}
\text{Tr}(\hat{D}^{\dagger}(n',\overline{x'},m',\overline{p'})\hat{D}(n,\overline{x},m,\overline{p}))=\delta(\overline{x}-\overline{x'})\delta(\overline{p}-\overline{p'}) \\ \cross \delta_{n',n}\delta_{m',m},
\end{multline}
where we used the formula $\sum_{n\in\mathds{Z}^{2}}e^{in(\overline{x}-\overline{x'})}=\delta(\overline{x}-\overline{x'})$ and satisfy the completeness relation:
\begin{equation}
\sum_{n,m\in\mathds{Z}^{2}}\int_{-l/2}^{l/2} \text{d}\overline{x} \int_{-\frac{\pi}{l}}^{\frac{\pi}{l}}  \text{d}\overline{p} \hat{D}(n,\overline{x},m,\overline{p}) \hat{D}^{\dagger}(n,\overline{x},m,\overline{p})=\mathds{I}
\end{equation}
Any operator can be expanded into such complete basis. In particular the density matrix reads:
\begin{equation}
\hat{\rho}=\int_{\frac{-l}{2}}^{\frac{l}{2}}\int_{-\frac{\pi}{l}}^{\frac{\pi}{l}}  \sum_{n,m\in\mathds{Z}^{2}} \hat{D}(n,\overline{x},m,\overline{p}) \chi(n,\overline{x},m,\overline{p}) \text{d}\overline{x} \text{d}\overline{p},
\end{equation}
where we introduced  the modular characteristic function $\chi(n,\overline{x},m,\overline{p})=\text{Tr}(\hat{\rho}\hat{D}^{\dagger}(n,\overline{x},m,\overline{p}))$.

\subsubsection{Point operator on a double cylinder phase space}
In this section, we introduce the point operator on a double cylinder phase space from its definition on the one cylinder phase space. It is defined as the symplectic Fourier transform of the displacement operator \cite{bizarro,common,rigas},
\begin{multline}\label{pointoperatordef}
\hat{\Delta}(n,m,\overline{x},\overline{p})=\int_{-\frac{\pi}{l}}^{\frac{\pi}{l}}\int_{-l/2}^{l/2} \text{d}\overline{x'}\text{d}\overline{p'}\sum_{q,s\in\mathds{Z}^{2}}e^{i\frac{2\pi q}{l}\overline{x}}e^{is\overline{p}l}\\ \cross e^{-i\frac{2\pi n}{l}\overline{x'}}e^{-im\overline{p'}l} \hat{D}(q,\overline{x'},s,\overline{p'}).
\end{multline}
An operator which has an expansion into a product of polynomial of $\hat{\overline{x}},\hat{\overline{p}},\hat{N},\hat{M}$ can be written in many ways depending on the choice of the  operators ordering. In Eq. (\ref{pointoperatordef}), the Weyl ordering has been considered \cite{ordering}. The variables of the point operator label the phase space and do not correspond to the variables of the integer basis.  
The point operator verifies the finite norm condition:
\begin{equation}
\sum_{n,m\in\mathds{Z}^{2}}\int_{-\frac{\pi}{l}}^{\frac{\pi}{l}}\int_{-l/2}^{l/2} \text{d}\overline{x}\text{d}\overline{p} \hat{\Delta}(n,m,\overline{x},\overline{p})=\mathds{I}
\end{equation}
Point and parity operators are related by the equation
\begin{equation}\label{equalitydis}
\hat{\Delta}(0,0,0,0)=\hat{\Delta}=\hat{\Pi},
\end{equation}
$\hat{\Pi}$ being the parity operator defined by $\hat{\Pi}=\int \text{d}x \ket{x}\bra{-x}$. $\hat{\Pi}$ is a reflection along the $x$-axis  and can also be written in the modular basis:
\begin{equation}\label{Parity}
\hat{\Pi}=\int_{-\frac{\pi}{l}}^{\frac{\pi}{l}}\int_{-l/2}^{l/2} \text{d}\overline{x}\text{d}\overline{p} \ket{\overline{x},\overline{p}} \bra{-\overline{x},-\overline{p}},
\end{equation}
where we used Eq.~(\ref{Zakt}). The parity operator can also be seen as an inversion about the origin in the modular plane. In the integer representation, the parity operator has a similar form, $\hat{\Pi}=\sum_{n,m\in\mathds{Z}^{2}}\ket{n,m}\bra{-n,-m}$.

We point out that the point operator $\hat{\Delta}(n,m,\overline{x},\overline{p})$ is also separable in the two pairs of azimuth-angular momentum variables, owing to the commutation relation invoked in \ref{Modularbasis}:
\begin{equation}
\hat{\Delta}(n,m,\overline{x},\overline{p})=\hat{f}(n,\overline{x})\hat{g}(m,\overline{p})
\end{equation}
which can be written, using Eq. (\ref{displacementinteger}), in the integer basis as:
{\small{\begin{multline}
\hat{f}(n,\overline{x})\hat{g}(m,\overline{p})=\sum_{n',r,m',r'\in\mathds{Z}}\int_{-l/2}^{l/2}\int_{-\pi/l}^{\pi/l} \text{d}\overline{x'} \text{d}\overline{p}' e^{i\overline{x'}\frac{2\pi}{l}(r+n'/2)}\\ \cross e^{-i(n'\overline{x}-n\overline{x'})\frac{2\pi}{l}}
e^{-i\overline{p'}l(r'+m'/2)}e^{i(m'\overline{p}-m\overline{p'}l)}\\ \cross  \ket{r+n',r'+m'}\bra{r,r'} .
\end{multline}}}
After integration, we obtain:
{\small{\begin{multline}
\hat{f}(n,\overline{x})\hat{g}(m,\overline{p})=\sum_{n',m'\in\mathds{Z}^{2}}e^{\frac{2i\pi}{l}\overline{x}(n-n'/2)}e^{i\overline{p}(m-m'/2)l} \\ \cross \ket{n+\frac{n'}{2},m+\frac{n'}{2}}\bra{n-\frac{m'}{2},m-\frac{m'}{2}}.
\end{multline}}}
Then, we can express the point operator under the covariant form:
\begin{equation}
\hat{\Delta}(n,m,\overline{x},\overline{p})=\hat{D}(n,\overline{x},m,\overline{p})\hat{\Delta}\hat{D^{\dagger}}(n,\overline{x},m,\overline{p}).\\
\end{equation}
where $\hat{\Delta}=\hat{\Pi}$ from Eq.~(\ref{equalitydis}). \\

In summary, the phase space described in this section  is the cartesian product of two cylinders $(S^{1}\cross\mathds{Z})^{2}$, since we have two independent pairs of canonically conjugate azimuthal-angular variable.

\subsection{Modular Wigner distribution}

The expectation value of the point operator \cite{pointoperator1,paz,wooters}, also called the symbol of the density matrix, is the quasi-probability  distribution in a double cylinder phase space $(S^{1}\cross\mathds{Z})^{2}$ whose variables are noted $(n,\overline{x})$ and $(m,\overline{p})$. We called it the Modular Wigner distribution and is defined as \cite{rigas,bizarro}:
\begin{equation}
W_{\hat{\rho}}(n,m,\overline{x},\overline{p})=\text{Tr}(\hat{\rho}\hat{\Delta}(n,m,\overline{x},\overline{p}))
\end{equation}
Due to the hermiticity of the point operator $\hat{\Delta}(n,m,\overline{x},\overline{p})$ the Modular Wigner distribution is real.
After taking the trace operation, the distribution takes the familiar form in the modular representation,
\begin{multline}\label{Wignermodular}
W_{\hat{\rho}}(n,m,\overline{x},\overline{p})=\int_{-\frac{\pi}{l}}^{\frac{\pi}{l}}\int_{-l/2}^{l/2} e^{2i\frac{2\pi}{l}n\overline{x'}}e^{-2i\overline{p'}ml} \text{d}\overline{x'}\text{d}\overline{p'}
 \\\cross \langle \overline{\overline{x}-\overline{x'}},\overline{\overline{p}-\overline{p'}}| \hat{\rho}| \overline{\overline{x}+\overline{x'}},\overline{\overline{p}+\overline{p'}}\rangle \end{multline}
In the integer representation, the Wigner distribution has the form:
{\small{\begin{multline}
W_{\hat{\rho}}(n,m,\overline{x},\overline{p})=w_{\hat{\rho}}(2n,2m,\overline{x},\overline{p})\\+\frac{2}{\pi}\sum_{n',m'\in\mathds{Z}} \frac{(-1)^{n-n'}}{2n'+1-2n}\frac{(-1)^{m-m'}}{2m'+1-2m}w_{\hat{\rho}}(2n'+1,2m'+1,\overline{x},\overline{p})
\end{multline}}}
where, 
\begin{multline}\label{Wignermodularinteger}
w_{\hat{\rho}}(2n+\alpha, 2m+\beta,\overline{x},\overline{p})=\sum_{n',m'\in\mathds{Z}^{2}} e^{-i(2n'+\alpha)\frac{2\pi}{l}\overline{x}} e^{i(2m'+\beta)\overline{p}l}\\
\cross \bra{2n-n',2m-m'}\hat{\rho}\ket{2n+n'+\alpha,2m+m'+\beta},
\end{multline}
with $\alpha,\beta=0,1$. We also note that another definition of Wigner distribution was proposed in \cite{hope} for the number-phase Wigner distribution. It consists of a phase space with half integer values. The two cylinder phase space $(\overline{x},n)$ and $(\overline{p},m)$ are not coupled  if the modular wave function in Eq.~(\ref{modularwave}) or the integer one in Eq.~(\ref{integerwavefunction}) is separable.

Summing over all variables of the distribution, we find the correct normalization of the modular Wigner distribution:
\begin{equation}
\int_{-\frac{\pi}{l}}^{\frac{\pi}{l}}\int_{-l/2}^{l/2} \text{d}\overline{x}\text{d}\overline{p} \sum_{n,m\in\mathds{Z}^{2}} W_{\hat{\rho}}(n,m,\overline{x},\overline{p}) =1,
\end{equation}
since the density matrix $\hat{\rho}$ is normalized. \\

The marginals can be found by summing over different variables. By summing over $n$ and $m$,  we obtain the following marginal,
\begin{equation}
\sum_{n,m\in\mathds{Z}^{2}}W_{\hat{\rho}}(n,m,\overline{x},\overline{p})=\bra{\overline{x},\overline{p}}\hat{\rho}\ket{\overline{x},\overline{p}}
\end{equation}
where we used the series representation of the Dirac distribution: $\delta(\overline{x})=\sum_{k\in\mathds{Z}}e^{ik\overline{x}}$. It corresponds to what we expect for such probability distribution as it is the diagonal matrix element of the density matrix in the modular basis. The marginal obtained by integrating over the  modular  variables is :
\begin{multline}\label{integerproba}
\int_{-\frac{\pi}{l}}^{\frac{\pi}{l}}\int_{-l/2}^{l/2} \text{d}\overline{x}\text{d}\overline{p} W_{\hat{\rho}}(n,m,\overline{x},\overline{p})= \bra{n,m}\hat{\rho}\ket{n,m},
\end{multline}
and gives also the correct marginal. The integers variables of the integer wave function $\psi_{n,m}$ cannot be interpreted as the integers which label the lattices along the $x$ and $p$-axis. Thus, the integer variables of the Wigner distribution do not have a simple physical meaning, which is a consequence of an additional phase. Nevertheless, the modular variables on phase space can be related to the coordinates of the two position and momentum lattices.  \\ 
We can define two partial traces of the density matrix. One of them is obtained by summing over the integer $n$ and by integrating over the  position modular variable $\overline{x}$:  
\begin{multline}\label{partialtrace}
F_{\hat{\rho}}(m,\overline{p})=\sum_{n\in\mathds{Z}}\int_{-\frac{l}{2}}^{\frac{l}{2}} \text{d}\overline{x} W_{\hat{\rho}}(n,m,\overline{x},\overline{p})\\=\int_{-\frac{\pi}{l}}^{\frac{\pi}{l}}\int_{\frac{-l}{2}}^{\frac{l}{2}} \text{d}\overline{x}\text{d}\overline{p'}\bra{\overline{x},\overline{\overline{p}-\overline{p'}}}\hat{\rho}\ket{\overline{x},\overline{\overline{p}+\overline{p'}}}e^{-2i\overline{p'}ml}
\end{multline}
The other partial trace is obtained simply by exchanging the pairs of variables $m$, $\overline{p}$ and $n$, $\overline{x}$. 
In addition, two crossed marginals can be determined by summing over $m$ (resp. $n$) and integrating over $\overline{p}$ (resp. $\overline{x}$):
\begin{align}
M_{1}(m,\overline{x}) =\bra{\overline{x},m}\hat{\rho} \ket{\overline{x},m}\\
M_{2}(n, \overline{p}) = \bra{\overline{p},n}\hat{\rho} \ket{\overline{p},n},
\end{align}
and they correspond to the correct marginals (see Eq.~(\ref{marginaleq})).
The density matrix can be obtained from the Modular Wigner distribution as:
\begin{equation}
\hat{\rho}=\sum_{n,m\in\mathds{Z}^{2}}\int_{-\frac{\pi}{l}}^{\frac{\pi}{l}}\int_{-l/2}^{l/2} \text{d}\overline{x} \text{d}\overline{p} \hat{\Delta}(n,m,\overline{x},\overline{p})W_{\hat{\rho}}(n,m,\overline{x}.\overline{p})
\end{equation}
With this last reconstruction property, the introduced distribution satisfies all the Stratanovich-Weyl postulates \cite{strato}.\\

We now explicitly study the case of two independent cylinder phase space, which corresponds to a separable bipartite system. For a pure state $\hat{\rho}=\ket{\psi}\bra{\psi}$, and a separable wave function $\psi(\overline{x},\overline{p})=f(\overline{x})g(\overline{p})$, we can write formally $\ket{\psi}= (\int \text{d}\overline{x} f(\overline{x}) \ket{\overline{x}})(\int \text{d}\overline{p} g(\overline{p}) )\ket{\overline{p}}=\ket{f}\ket{g}$.  Therefore, the modular Wigner distribution is also separable, and we can write:
\begin{equation}
W_{\hat{\rho}}(n,m,\overline{x},\overline{p})=W_{\ket{f}}(n,\overline{x})W_{\ket{g}}(m,\overline{p}).
\end{equation}
Each distribution is an angle-momentum Wigner distribution and the two cylinders phase space are hence decoupled.
In that case, the modular Wigner distribution in the integer representation takes the form:
\begin{align}\label{Wignerseparable}
W_{\ket{f}}(n,\overline{x})=\int_{-l/2}^{l/2} e^{2i\frac{2\pi}{l}n\overline{x'}} \langle \overline{\overline{x}-\overline{x'}}|\ket{f}\bra{f}|\overline{\overline{x}+\overline{x'}}\rangle\text{d}\overline{x'}\\
W_{\ket{g}}(m,\overline{p})=\int_{-\frac{\pi}{l}}^{\frac{\pi}{l}} e^{-2im\overline{p'}l} \langle \overline{\overline{p}-\overline{p'}}| \ket{g}\bra{g}| \overline{\overline{p}+\overline{p'}}\rangle \text{d}\overline{p'}
\end{align}
We note that the modular Wigner distribution $W_{\ket{g}}(m,\overline{p})$ has similar properties to the one introduced in Ref. \cite{rigas}. We also point out that $F(m,\overline{p})$, whose expression is given by Eq.~(\ref{partialtrace}), when the state is pure and $\ket{\psi}=\ket{f}\ket{g}$, is related to $W_{\ket{g}}(m,\overline{p})$ up to a multiplicative factor.

\section{Modular and integer representation of GKP state} \label{sectionfour}
In this section, we recall a few properties of GKP states \cite{GKP,Albert}, which are translational symmetric bosonic codes and express them using the modular representation.

\subsection{Ideal GKP states}
\begin{figure}[h!]
\begin{center}
 \includegraphics[scale=0.35]{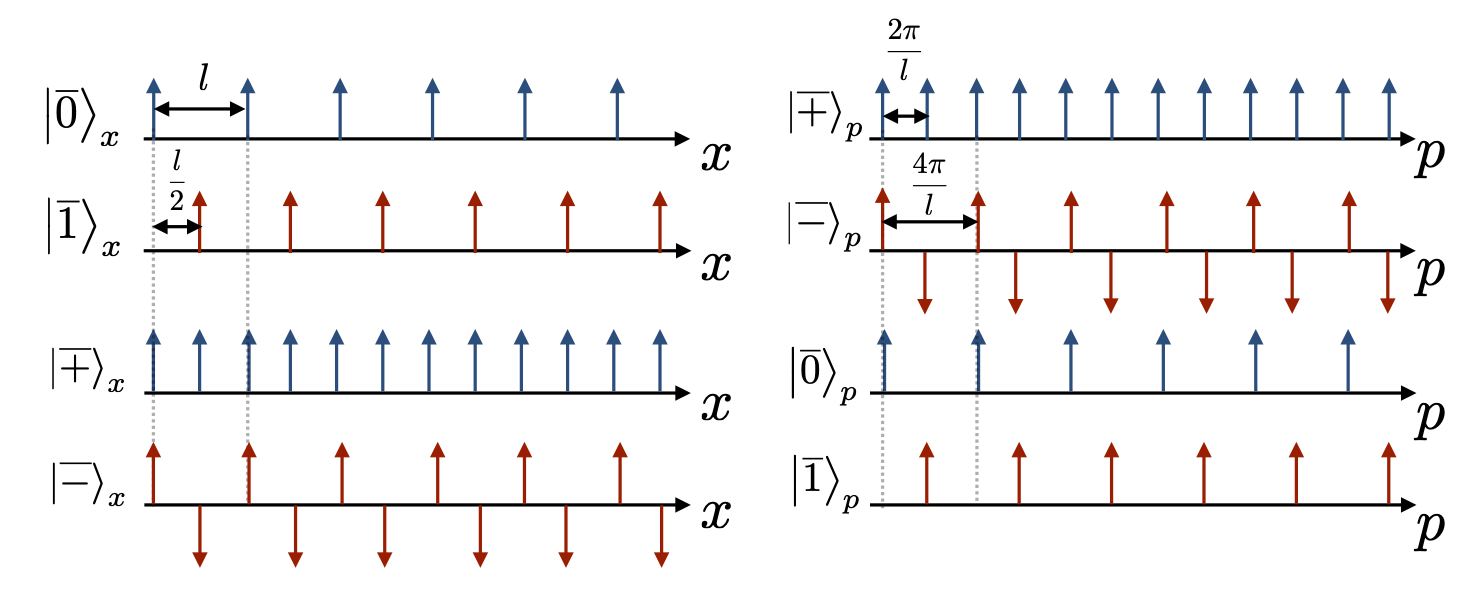}
 \caption{\label{GKPrep}Schematic plot of the wave function of the GKP state in the two orthogonal quadratures. The vertical arrows represent the Dirac comb. In the $x$-representation, the $\ket{\overline{0}}_{x}$ (in blue) and $\ket{\overline{1}}_{x}$ (in red) logical state has a $l$ periodicity and correspond to the $\ket{\overline{+}}_{p}$ and $\ket{\overline{-}}_{p}$ logical states in the $p$-representation. }
 \end{center}
\end{figure}

In order to define the GKP qubit using continuous variables, we start by defining a lattice along the $x$-axis with an interval length $l=\sqrt{\pi}$. This corresponds to a lattice along the $p$-axis of length $2\sqrt{\pi}=2\pi/\sqrt{\pi}$ as in the case of modular variables. The $\ket{\overline{0}}$ logical state of the qubit is defined as the translationally invariant comb with period $2\sqrt{\pi}$, where each peak is centered in each interval. The $\ket{\overline{1}}$ logical state is defined by the translation of the  $\ket{\overline{0}}$ logical state of a $\sqrt{\pi}$ length. The GKP states are non-physical, since they are composed of infinitively squeezed peaks with an infinite large envelope as represented on Fig.~\ref{GKPrep}. The two logical states can be written as  two localized states in the modular plane:
\begin{align}\label{IdealGKP}
\ket{\overline{0}}_{x}=|\overline{x}=-\frac{l}{4},\overline{p}=0\rangle=\sum_{n \in \mathds{Z}} | -\frac{l}{4}+nl\rangle,\\
\ket{\overline{1}}_{x}=|\overline{x}=\frac{l}{4},\overline{p}=0\rangle=\sum_{n \in \mathds{Z}} | \frac{l}{4}+nl \rangle.
\end{align}
It explicits the interest of using the modular basis to describe these states. The $\ket{\overline{0}}$ (resp. $\ket{\overline{1}}$) logical state is a point centered at $\mp l /4$ on the left (resp. right) modular half-plane (see Fig.~(\ref{latticevariablemodular})).
We also define the linear superposition $\ket{\overline{\pm}}_{x}=\frac{1}{\sqrt{2}}(\ket{\overline{0}}_{x}\pm \ket{\overline{1}}_{x})=\ket{\overline{0},\overline{1}}_{p}$ where the last equality stands only if the periodicity of the GKP states is $\sqrt{\pi}$.  \\

In the integer basis, the GKP states are delocalized. This can be intuitively understood, since the modular and the integer basis are related by a Fourier series transform and is written as:  
\begin{align}\label{gkpinteger}
\ket{\overline{0}}_{x}=\sum_{n,m\in\mathds{Z}^{2}} e^{-i\pi n/2}\ket{n,m}\\
\ket{\overline{1}}_{x}=\sum_{n,m\in\mathds{Z}^{2}} e^{i\pi n/2}\ket{n,m}
\end{align}
 The fact that the integer wave function does not depend on $m$ is a consequence of the non-normalizability of the state. It will be interpreted in the next section.\\

Alternatively we can define the lattice along the $p$-quadrature exchanging the definition of  $\ket{\overline{0}}_{p}$ by $\ket{\overline{+}}_{p}$. This remark will be useful for Sec. \ref{idealmodularwigner}.

\subsection{Construction of physical GKP state}
In this section, we describe a physical GKP state using the formalism developed in \cite{Glancy,motes}. For such, we will consider noise model in the integer and modular variables rather than in the position and momentum quadrature variables, which are equivalent in the limit described below. \\

 We start by considering small shifts in position ${\cal{\hat{D}}}(u,0)=e^{iu\hat{p}}$ and momentum ${\cal{\hat{D}}}(0,v)=e^{-iv\hat{q}}$ quadrature where $u\in [-\sqrt{\pi}/2,\sqrt{\pi}/2[$ and $v\in [-\sqrt{\pi},\sqrt{\pi}[$ applied to an ideal GKP state. The application of such small displacements to the state $\ket{\overline{0}}_{x}$ can be written in the modular basis:
\begin{equation}\label{idealnoisystate}
e^{-iv\hat{q}}e^{iu\hat{p}}\ket{\overline{0}}_{x}=\ket{\overline{x}=u,\overline{p}=v}
\end{equation}
If the displacement goes beyond the modular plane $u,v>\sqrt{\pi}$, additional phase factors appear. In addition, for small displacements, the displacement operator ${\cal{\hat{D}}}(u,0)$ and ${\cal{\hat{D}}}(0,v)$ coincide with the modular displacement operator $\hat{D}(0,\overline{x}=u)$ and $\hat{D}(0,\overline{p}=v)$. Hence, the noise models, using position-momentum variables or the bounded/integer position and momentum one,  are equivalent in the limit of small shifts {\it{i.e}} $\abs{u},\abs{v}<\sqrt{\pi}/6$ \cite{Glancy}. It will be useful for finding a criteria of the correctability of GKP state as developed in \cite{Glancy} and in Sec. \ref{gkpcorrectionsection}.

 Physical GKP states are constructed by applying the operator $\hat{U}$, expanded in the modular displacement basis, to the ideal GKP state $\ket{\tilde{0}}_{x}=\hat{U}\ket{\overline{0}}_{x}$ defined by:
\begin{multline}\label{logicalconstruction}
\ket{\tilde{0}}_{x}= \int_{-\frac{\pi}{l}}^{\frac{\pi}{l}}\int_{\frac{-l}{2}}^{\frac{l}{2}} \sum_{n,m\in\mathds{Z}^{2}} E(\overline{x},n, \overline{p}, m) \hat{D}(\overline{x},n)\hat{D}(\overline{p},m) \\ \text{d}\overline{x} \text{d}\overline{p} \ket{\overline{0}}_{x},
\end{multline}
where $E(\overline{x},n, \overline{p}, m)$ is the noise distribution or the Weyl-operator expansion of $\hat{U}$. This procedure for building physical GKP states is analog to the one described in \cite{motes}. \\
After applying the displacement operators to the ideal GKP state, we get:
\begin{equation}
\hat{D}(\overline{x},n)\hat{D}(\overline{p},m) \ket{\overline{0}}_{x}=e^{-2i\pi\overline{x}n/l}e^{-i\pi n/2}e^{-i\overline{p}ml}\ket{\overline{x}-l/4,\overline{p}} 
\end{equation}
Using the previous equation, the state described by Eq.~(\ref{logicalconstruction}) can be written as:

{\small{\begin{multline}
\ket{\tilde{0}}_{x}=\int_{-\frac{\pi}{l}}^{\frac{\pi}{l}}\int_{\frac{-l}{2}}^{\frac{l}{2}} \sum_{n,m\in\mathds{Z}^{2}} E(\overline{x},n, \overline{p}, m) e^{-2i\pi\overline{x}n/l}e^{-i\pi n/2}e^{-i\overline{p}ml}  \\\cross \ket{\overline{x}-l/4,\overline{p}} \text{d}\overline{x} \text{d}\overline{p}.
\end{multline}}}
After performing a change of variable on $\overline{x}$, the real GKP state reads:
\begin{equation}\label{real0}
\ket{\tilde{0}}_{x}=\int_{-\frac{\pi}{l}}^{\frac{\pi}{l}}\text{d}\overline{p} \int_{\frac{-l}{2}}^{\frac{l}{2}} \text{d}\overline{x} \psi(\overline{x},\overline{p}) \ket{\overline{x},\overline{p}},
\end{equation}
where the modular wavefunction is $\psi(\overline{x},\overline{p}) = \sum_{n,m} E(\overline{x},n, \overline{p}, m)e^{-2i\pi\overline{x}n/l}e^{-i\pi n/2}e^{-i\overline{p}ml}$.\\

Square GKP states, whose envelope and peaks of the comb are Gaussian functions \cite{GKP,Glancy,Albert}, can be obtained assuming that the noise distribution $E$ is separable in the pairs of variables: $E(\overline{x},n, \overline{p}, m)=f(\overline{x},n)h(\overline{p},m)$. Then, to obtain such square GKP states, the sums over $n$ and $m$ must be equal to two Gaussian combs on each variable, namely $\sum_{n}e^{-2i\pi\overline{x}n/l}e^{-i\pi n/2}f(\overline{x},n) \cross \sum_{m} e^{-i\overline{p}ml} h(\overline{p},m)=\sum_{n}G_{\Delta}(\overline{x}-nl)\sum_{m}G_{\kappa}(\overline{p}-ml)$, where $G$ denotes the Gaussian function: $G_{\Delta}(\overline{x})=\text{exp}(-\overline{x}^{2}/(2\Delta^{2}))$. For such, we can use the Poisson summation formula to specify $f$ and $g$ : $\sum_{n\in\mathds{Z}} s(t+nl)=\sum_{k\in\mathds{Z}} \tilde{s}(k/l) e^{2i\pi kt/l}$, where $\tilde{s}$ is the Fourier transform of $s$. We obtain:
\begin{align}
f(\overline{x},n)=e^{-(n\Delta/l)^{2}/2},\\
h(\overline{p},m)=e^{-(ml\kappa/2\pi)^{2}/2},
\end{align}
which do not depend explicitly on the modular variables.

Finally, the modular wave function $\psi(\overline{x},\overline{p})$ of square GKP states can be written as: 
\begin{equation}\label{GaussianGKP}
\psi(\overline{x},\overline{p}) = N(l,\kappa,\Delta)\cdot G_{\Delta}(\overline{x}-l/4)G_{\kappa}(\overline{p}),
\end{equation}
where $N(l,\kappa,\Delta)=1/(2.\text{erf}(\frac{\kappa\pi}{l}).(\text{erf}(\frac{l}{4\Delta})+\text{erf}(\frac{3l}{4\Delta}))^{1/2}$ is a normalization constant and depends on the lattice lenght $l$. The function erf is the error function defined by $\text{erf}(l/2\Delta)=\int_{-l/2}^{l/2}e^{-\overline{x}^{2}/2\Delta^{2}} \text{d}\overline{x}$. The modular wave function of the $\ket{\tilde{1}}_{x}$ state can be written as $\psi(\overline{x},\overline{p})=N(l/\sigma).G_{\Delta}(\overline{x}+l/4)G_{\kappa}(\overline{p})$, with the same normalisation factor. \\

\begin{figure}[h!]
\begin{center} \includegraphics[scale=0.35]{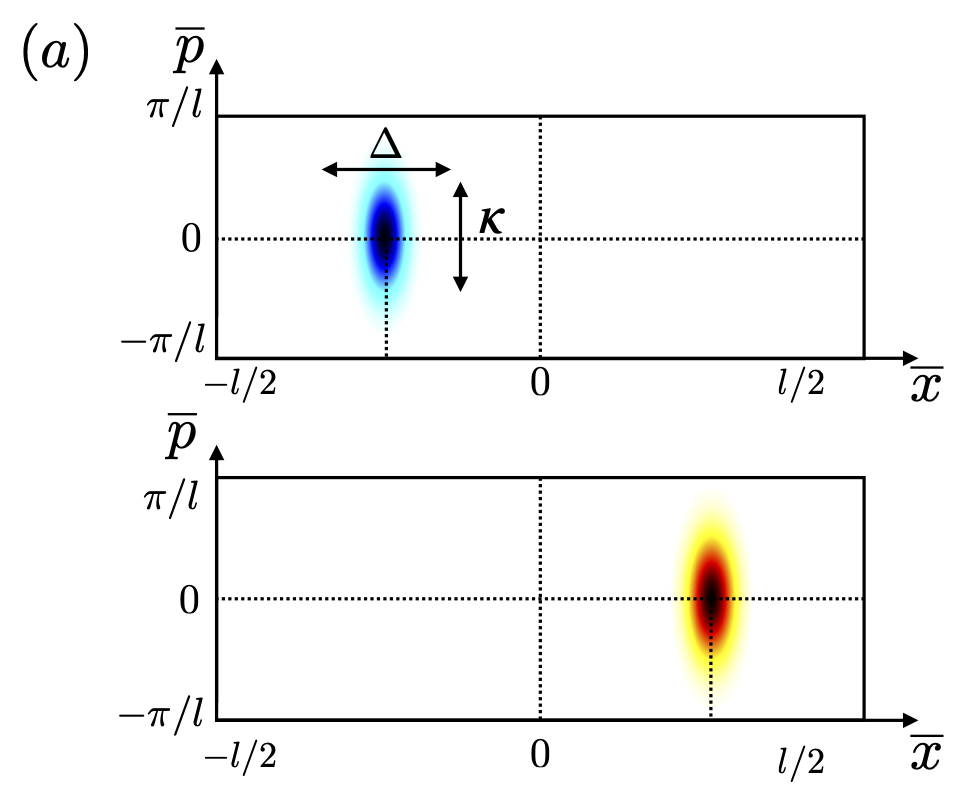}
 \includegraphics[scale=0.35]{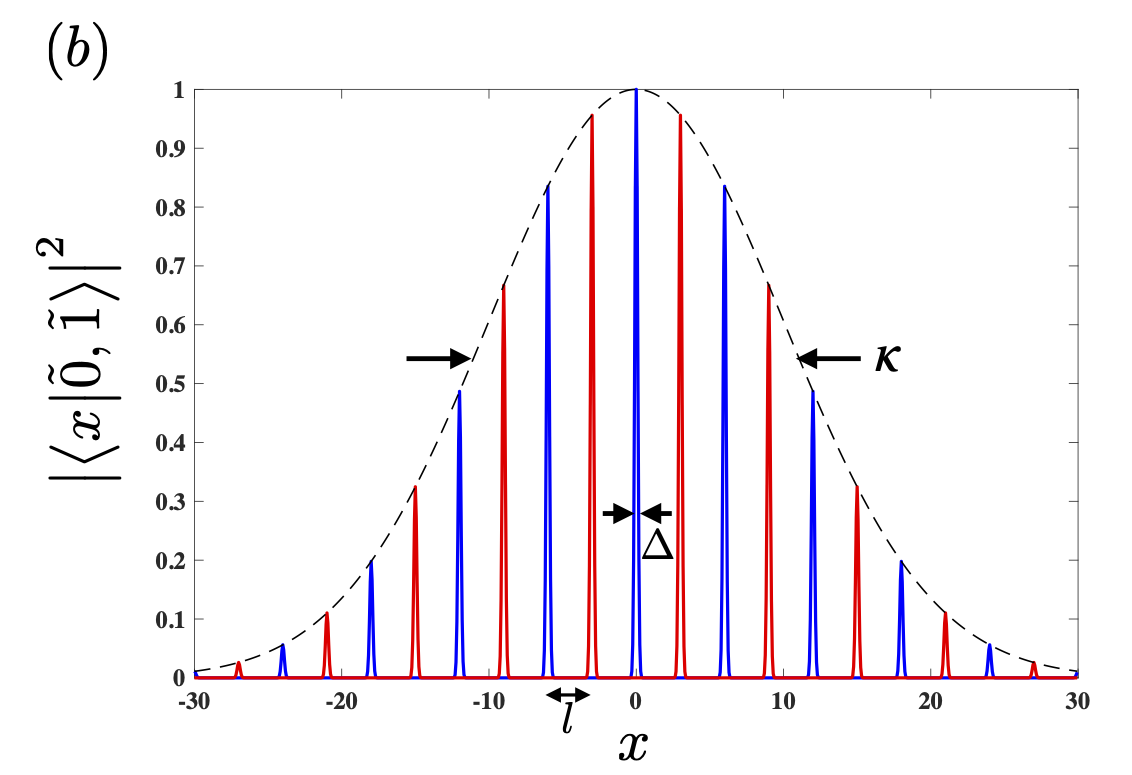}
 \caption{\label{GKPfigure} (a) Probability density of the $0$ (blue) and $1$ (red) physical GKP state in the modular plane, which are Gaussian in the variable $\overline{x}$ and $\overline{p}$. (b) Wave function of the GKP state in the $x$-representation. }
 \end{center}
\end{figure}

The real GKP states are represented  on Fig.~(\ref{GKPfigure})(a) and (b) in the modular plane (resp. over the real line $x$)  when $l \gg \Delta$ and $ 2\pi/l \gg \kappa$, which are the conditions assumed here. \\

The scalar product between the two logical states is:
\begin{equation}\label{scalarproduct}
 \bra{\tilde{0}}\ket{\tilde{1}}_{x}=\frac{\text{erf}(\frac{\kappa\pi}{l}).e^{-(l/4\Delta)^{2}}\text{erf}(l/2\Delta)}{\text{erf}(3l/4\Delta)+\text{erf}(l/4\Delta)}
 \end{equation}
The two states become orthogonal in the limit $l\gg \Delta$. The two logical states are represented in Fig.~\ref{GKPfigure} in the modular plane as well as a function of $x$. Finally, following Ref. \cite{Glancy}, the modular wave function for GKP states can be interpreted as the amplitude of probability of having an error of $u$ (resp. $v$) in the $x$-quadrature (resp. $p$) if $\abs{u},\abs{v}<\sqrt{\pi}/6$.

The state described by Eq.~(\ref{real0}) can be written in the integer representation: $\ket{\tilde{0}}=\sum_{n,m \in \mathds{Z}^{2}}f_{n}g_{m}\ket{n,m}$ with $f_{n}=\int_{-l/2}^{l/2} \text{d}\overline{x} e^{-i\frac{2\pi}{l}n\overline{x}}G_{\Delta}(\overline{x}+l/4)$ and $g_{m}=\int_{-\pi/l}^{\pi/l} \text{d}\overline{p} G_{\kappa}(\overline{p}) e^{im\overline{p}l}$ that are calculated in details in Appendix \ref{GKPinteger}. When we consider ideal GKP states (see Eq. (\ref{gkpinteger})), the envelop of the position comb is infinite and this is why $g_{m}$ does not depend of $m$.

As in the ideal case, we can also define the linear superposition of the two physical logical GKP state,
\begin{multline}\label{gaussian+}
\ket{\tilde{\pm}}_{x}=\frac{\tilde{N}(l,\kappa,\Delta)}{\sqrt{2}}(\int_{-\frac{\pi}{l}}^{\frac{\pi}{l}}\text{d}\overline{p} \int_{\frac{-l}{2}}^{\frac{l}{2}} \text{d}\overline{x} (G_{\Delta}(\overline{x}-l/4)G_{\kappa}(\overline{p})\\ \pm G_{\Delta}(\overline{x}+l/4)G_{\kappa}(\overline{p}))\ket{\overline{x},\overline{p}}), 
\end{multline}
 where the non-trivial normalization $\tilde{N}(l/\sigma)=1/(\abs{\bra{\tilde{0}}\ket{\tilde{0}}+\bra{\tilde{1}}\ket{\tilde{1}}\pm2\bra{\tilde{0}}\ket{\tilde{1}}}^{1/2})$ with $\bra{\tilde{0}}\ket{\tilde{0}}=\bra{\tilde{1}}\ket{\tilde{1}}=2\text{erf}(\frac{\kappa\pi}{l})\cdot(\text{erf}(\frac{l}{4\Delta})+\text{erf}(\frac{3l}{4\Delta}))$ and $\bra{\tilde{0}}\ket{\tilde{1}}$ is given by Eq.~(\ref{scalarproduct}).
This formalism could be used for building hexagonal GKP state \cite{Albert}, and the corresponding functions $f$ and $h$ could be determined with a numerical approach.

\subsection{Set of logical of gates}

We now introduce the set of Pauli matrices adapted for this qubit encoding \cite{Keterrermodular}. It corresponds to the displacement operators:
\begin{align}
\hat{X}={\cal{\hat{D}}}(\frac{l}{2},0)= \int_{-l/2}^{l/2}  \text{d}\overline{x} \int_{-\pi/l}^{\pi/l} \text{d}\overline{p} e^{2i\pi\overline{x}/l}  \hat{\sigma}_{x}(\overline{x},\overline{p})
\end{align}
\begin{equation}
\hat{Z}={\cal{\hat{D}}}(0,\pi/l)= \int_{-l/2}^{l/2}  \text{d}\overline{x} \int_{-\pi/l}^{\pi/l} \text{d}\overline{p} e^{-i\overline{p}l} \hat{\sigma}_{z}(\overline{x},\overline{p})
\end{equation}
with the Pauli matrices
 \begin{equation}
 \hat{\sigma}_{x}(\overline{x},\overline{p})=e^{-i\overline{p}l/2}\ket{\overline{x},\overline{p}}\langle \overline{x}+\frac{l}{2},\overline{p}|+e^{i\overline{p}l/2} |\overline{x}+\frac{l}{2},\overline{p}\rangle \bra{\overline{x},\overline{p}}
 \end{equation}
 \begin{equation}
 \hat{\sigma}_{y}(\overline{x},\overline{p})=i(e^{i\overline{p}l/2}| \overline{x}+\frac{l}{2},\overline{p}\rangle \bra{\overline{x},\overline{p}}-e^{-i\overline{p}l/2}\ket{\overline{x},\overline{p}}\langle \overline{x}+\frac{l}{2},\overline{p}|)
 \end{equation}
  \begin{equation}\label{sigmaz}
 \hat{\sigma}_{z}(\overline{x},\overline{p})=\ket{\overline{x},\overline{p}}\bra{\overline{x},\overline{p}}-|\overline{x}+\frac{l}{2},\overline{p}\rangle \langle \overline{x}+\frac{l}{2},\overline{p}|
 \end{equation}
These gates are not hermitian and that $\hat{Z}^{2} \neq \mathds{I}$ and $\hat{X}^{2}\neq\mathds{I}$. Operators $\hat{X}$ and $\hat{Z}$ are not elements of the basis of the $\mathds{SU}(2)$ Lie algebra.  In \cite{Keterrermodular}, it was shown that we can introduce modular readout observables $\hat{\Gamma}_{\beta}$, which are hermitian  and constructed by analogy with the logical Pauli operators. They are defined by:
\begin{equation}\label{modulareadout}
\hat{\Gamma}_{\beta}=\int_{-l/2}^{l/2} \text{d}\overline{x}\int_{-\pi/l}^{\pi/l} \text{d}\overline{p} \hat{\sigma}_{\beta}(\overline{x},\overline{p}), 
\end{equation}
 with $\beta=x,y,z$ and with the modular Pauli matrices $ \hat{\sigma}_{\beta}(\overline{x},\overline{p})$. These operators will be useful for the tomographic reconstruction of the modular Wigner distribution in Sec.\ref{sectionseven}.

\section{Examples of Modular Wigner distribution }\label{sectionfive}
In this section, we discuss the double cylinder phase space representation of GKP states and Gaussian states.

\subsection{Modular Wigner distribution of GKP state}

\subsubsection{Ideal GKP state}\label{idealmodularwigner}
The modular wave function of the GKP states described by Eq.~(\ref{IdealGKP}) is $\psi(\overline{x},\overline{p})=\delta(\overline{x}\pm l/4)\delta(\overline{p})=f(\overline{x})g(\overline{p})$ with $l=\sqrt{\pi}$ and has a corresponding modular Wigner distribution  $W_{\ket{\overline{0},\overline{1}}_{x}}(n,m,\overline{x},\overline{p})=W_{\ket{f}}(n,\overline{x})W_{\ket{g}}(m,\overline{p})$: 
\begin{align}
W_{\ket{f}}(n,\overline{x})=\delta(\overline{x}\pm l/4),\\
W_{\ket{g}}(m,\overline{p})=\delta(\overline{p}),
\end{align}
and do not depend on $n$ and $m$ owing to the non-normalizability of the ideal GKP states as mentioned in \cite{rigas}. Indeed, a localized state with an associated modular variable Dirac distribution is constant in the integer (canonical conjugate) basis.   For a coherent superposition of the qubit state $\ket{\overline{\pm}}_{x}=\frac{1}{\sqrt{2}}(\ket{\overline{0}}_{x}\pm\ket{\overline{1}}_{x})$ with associated modular wave function $\psi(\overline{x},\overline{p})=\frac{1}{\sqrt{2}}(\delta(\overline{x}-l/4)+\delta(\overline{x}+l/4))\delta(\overline{p})$, the modular Wigner distribution $W_{\ket{\overline{\pm}}_{x}}(n,m,\overline{x},\overline{p})$ is given by:
\begin{align}\label{wignergkp}
&W_{\ket{f}}(n,\overline{x})=\delta(\overline{x}-l/4)+\delta(\overline{x}+l/4)\pm2\delta(\overline{x})\text{cos}(\frac{n\pi}{\hbar})\\
&W_{\ket{g}}(m,\overline{p})=\delta(\overline{p}).
\end{align}
The argument of the cosinus function do not depend on $l$ since the state is localized in $\pm l/4$ and $n$ is a multiple of $2\pi/l$. It only depends on $\pi/\hbar$, where $\pi$ has the dimension of an action. The distribution of this state is represented in Fig.~\ref{wignermodular}. The modular Wigner distributions of the coherent superposition of GKP states $\ket{\overline{\pm}}$ have the shape of ideal even/odd Schrödinger cat states (with an infinite squeezing along one quadrature), and a periodicity of the fringes equals to the distance of the two localized state. The shape of such state in cylinder phase space can be easily understood since the GKP state $\ket{\overline{\pm}}_{x}$ are superposition of localized states in the torus.\\

The presence of oscillations in the cylinder phase space $(n,\overline{x})$ (see Eq. ~(\ref{wignergkp})) and not in the one $(m,\overline{p})$ one is not a special feature of this phase space but rather a matter of definition of the GKP states. Indeed, exchanging the definition of $\ket{\overline{0}}_{p}$ and $\ket{\overline{+}}_{p}$ (the two states being identical), the oscillations from the coherent superposition would appear on the other cylinder $(m,\overline{p})$.\\

We now study the effect of small shifts on the previous states $\ket{u,v}=\hat{D}(u,0))\hat{D}(v,0)\ket{\overline{\pm}}_{x}$ where $l=\sqrt{\pi}\gg u,v$ corresponds to a displaced ideal cat states in the modular plane and are not localized at the center of the left and right modular plane. The associated modular Wigner distribution $W_{\ket{f,g}}(n,m,\overline{x},\overline{p})$ is:
{\small{\begin{align}
&W_{\ket{f}}(n,\overline{x})=\delta(\overline{x}-\overline{\frac{l}{4}-u})+\delta(\overline{x}+\overline{\frac{l}{4}-u})\pm2\delta(\overline{x})
\text{cos}(\frac{n\pi}{\hbar})\\
&W_{\ket{g}}(m,\overline{p})=\delta(\overline{p}+v).
\end{align}}}
These displacements can be induced by noise. They shift the full Modular Wigner distribution in the two uncoupled cylinders phase space. \\

\subsubsection{Physical GKP states}

\begin{figure}[h!]
\begin{center}
 \includegraphics[scale=0.45]{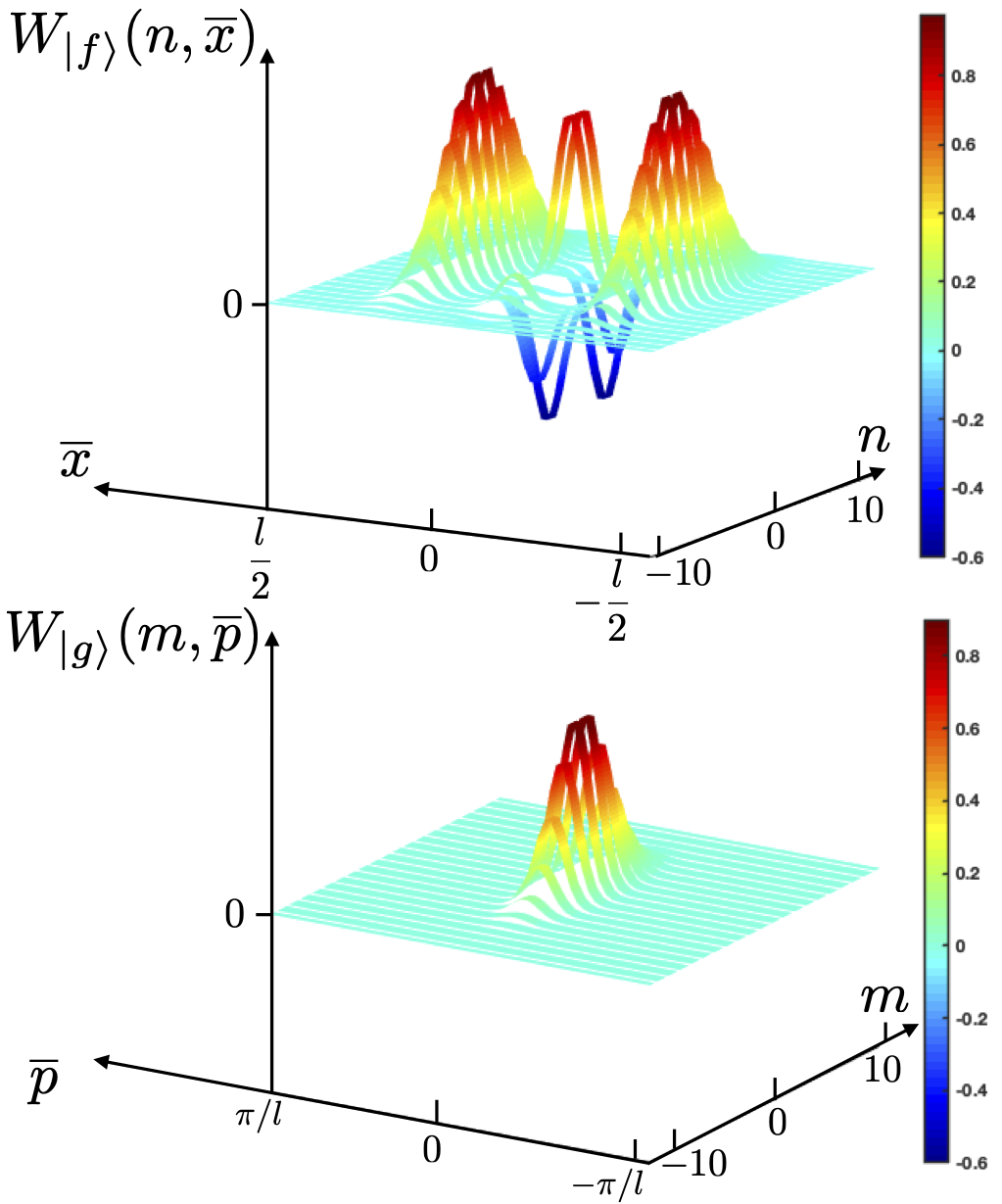}
 \caption{\label{wignermodular}Modular Wigner distribution of the physical GKP state for state with $l\gg \Delta$ which have the aspect of a Schrödinger cat state in the $(\overline{x},n)$ cylinder phase space. The two parameters are here set to $\Delta=0.1$ and $l=1$. }
 \end{center}
\end{figure}

We now consider  physical GKP states which are not translationally invariant due to the finite envelope in both modular variables (see Eq.~(\ref{GaussianGKP}) and Fig.~\ref{GKPfigure}). Using the condition $l\gg \Delta$, it is possible to compute the analytical expression of the modular Wigner distribution of the physical GKP state corresponding to the Gaussian modular wave function Eq.~(\ref{gaussian+}). The wave function being separable in $(\overline{x},\overline{p})$,  the modular Wigner distribution $W_{\ket{\tilde{\pm}}}(n,m,\overline{x},\overline{p})$ is  also separable into two parts, the interference part:
{\small{\begin{multline}\label{maingkp}
W_{\ket{f}}(n,\overline{x})\simeq e^{-\frac{(\overline{x}+l/4)^{2}}{2\Delta^{2}}}e^{-n^{2}(2\pi\Delta/l)^{2}}+e^{-\frac{(\overline{x}-l/4)^{2}}{2\Delta^{2}}}e^{-n^{2}(2\pi\Delta/l)^{2}}\\+2.e^{-n^{2}(2\pi\Delta/l)^{2}}e^{-\frac{\overline{x}^{2}}{\Delta^{2}}}\text{cos}(\frac{n\pi}{\hbar}),
\end{multline}}}
 and the envelope part:
\begin{equation}\label{envelopgkp}
W_{\ket{g}}(m,\overline{p})=\text{exp}(-\overline{p}^{2}\kappa^{2})\text{exp}(-m^{2}l^{2}/\kappa^{2}).
\end{equation}
The modular Wigner distribution in the two cylinder phase space is shown on Fig.~\ref{wignermodular}. The visibility of the oscillation is one since the state is a pure state but the purity $P=\text{Tr}(\hat{\rho}^{2})$ of the GKP states could be reduced if the state crosses a Gaussian channel \cite{Glancy,Albert}.

The number of oscillations on the cylinder phase space can be related to the probability of having an error in the encoding and will be studied in Sec. \ref{sectionsix}.

\subsection{$\pi/2$ Rotation in the cylinder phase space of the GKP states}

In this section, we study the Modular Wigner function of the state $\psi_{\pi/2}$ obtained after a $\pi/2$ rotation in the two cylinder phase plane of the ideal GKP state $\ket{\overline{\pm}}_{x}$. In the $(n, \overline{x})$ cylinder phase space, the modular Wigner distribution of this $\pi/2$ rotated GKP state corresponds to  two peaks along the integer $n$-axis, and presents an oscillation pattern along the modular variable direction $\overline{x}$. In the other decoupled cylinder phase space $(m, \overline{p})$, the Dirac point along the $\overline{p}$ axis is after the $\pi/2$ rotation along the $m$ axis. This ideal state, noted $\ket{\psi_{\pi/2}}$, constitutes a generalization of the OAM (orbital angular momentum) state described in \cite{rigas}. This is a consequence of the double cylinder aspect considered in this paper. The state is a coherent superposition in the integers basis and can be written under the form:
\begin{equation}\label{pirotated}
\ket{\psi_{\pi/2}}=\frac{1}{\sqrt{2}}(\ket{n_{0},m_{0}}\pm \ket{-n_{0},m_{0}})
\end{equation}
where $n_{0}$ is an even integer, and $m_{0}$ can be any integer. Using Eq.~(\ref{Wignermodularinteger}), the corresponding modular Wigner distribution $W_{\ket{\psi_{\pi/2}}}(n,m,\overline{x},\overline{p})=W_{\ket{f}}(n,\overline{x})W_{\ket{g}}(m,\overline{p})$ is 
{\small{\begin{align}
W_{\ket{f}}(n,\overline{x})=\delta_{n,n_{0}}+\delta_{n,-n_{0}}\pm\delta_{n} \cross \text{cos}(\frac{8\pi\overline{x}}{\hbar l}),\\
W_{\ket{g}}(m,\overline{p})=\delta_{m,m_{0}}.
\end{align}}}
As in Eq.~(\ref{wignergkp}), the Modular Wigner distribution is flat in the modular variables $\overline{x}$ and $\overline{p}$ due to the non-normalizability of the state described by Eq.~(\ref{pirotated}). Calculating the modular Wigner distribution of the corresponding physical state of $\ket{\psi_{\pi/2}}$ using a Gaussian wave function instead of Dirac one could be handled using results in \cite{Hinarejos}. We can now ask the question of how to perform such a $\pi/2$ rotation in the two uncoupled cylinders phase plane of a GKP states. A way to do so is to apply the sum of two projectors on the state defined by Eq.~(\ref{gkpinteger}): $\frac{1}{\sqrt{2}}(\hat{\Pi}(n_{0},m_{0})\pm \hat{\Pi}(-n_{0},m_{0}))\ket{\overline{+}}_{x}\ket{\overline{+}}_{x}$, where $\hat{\Pi}(\pm n_{0},m_{0})=\ket{\pm n_{0}, m_{0}}\bra{n_{0},m_{0}}$. The state $\hat{\Pi}(n_{0},m_{0})$ has a similar structure of a quantum state of a particle that passed through a diffraction slit (see (Eq.~(\ref{integertox})). Hence, using the transversal degree of freedom of single photon, this operation could be physically implemented  with a spatial light modulator (SLM) \cite{SLM}. An alternative way would be to  use the  Fourier series relation between the modular basis and the integer one as in Eq.~(\ref{modtoint}) and Eq.~(\ref{inttomod}) and could be also implemented thanks to a SLM.

\subsection{Modular Wigner distribution of a coherent and cat state}
 Gaussian states  are by definition states whose Wigner distribution in rectangular phase space are Gaussians \cite{gaussianstate}. They play a fundamental role in CV quantum information since they can be easily produced experimentally, as coherent states and squeezed states. QC involving only Gaussian states and Gaussian operations can be efficiently simulated with a classical computer \cite{SpeedupCVQC}. Nevertheless, they constitute a building block for CV quantum computation, as they are needed for the elaboration of cubic phase gate \cite{cubic,cubic2} or as input of continuous variable algorithm such as Gaussian Boson sampling \cite{gbs,hafnians2}. A pure Gaussian state has always a positive Wigner distribution according to Hudson's theorem \cite{hudson}, and the negativity of the Wigner distribution in the rectangular phase plane is frequently associated to the quantumness of a state. In the cylinder phase plane, a Gaussian state can have a negative Wigner distribution. As a matter of fact in such phase space, the only state which has a positive Wigner distribution is the eigenvector of both $\hat{N}$ and $\hat{M}$, {\it{i.e}} the state $\ket{n,m}$, which can be seen as a direct generalization of the results on the single cylinder phase space \cite{common}.\\
 
  In this paragraph, we study the modular wave function of a coherent state and of a cat state (a non-Gaussian one) and calculate analytically their associated modular Wigner distribution.\\

\begin{figure}[h!]
\begin{center}
 \includegraphics[scale=0.45]{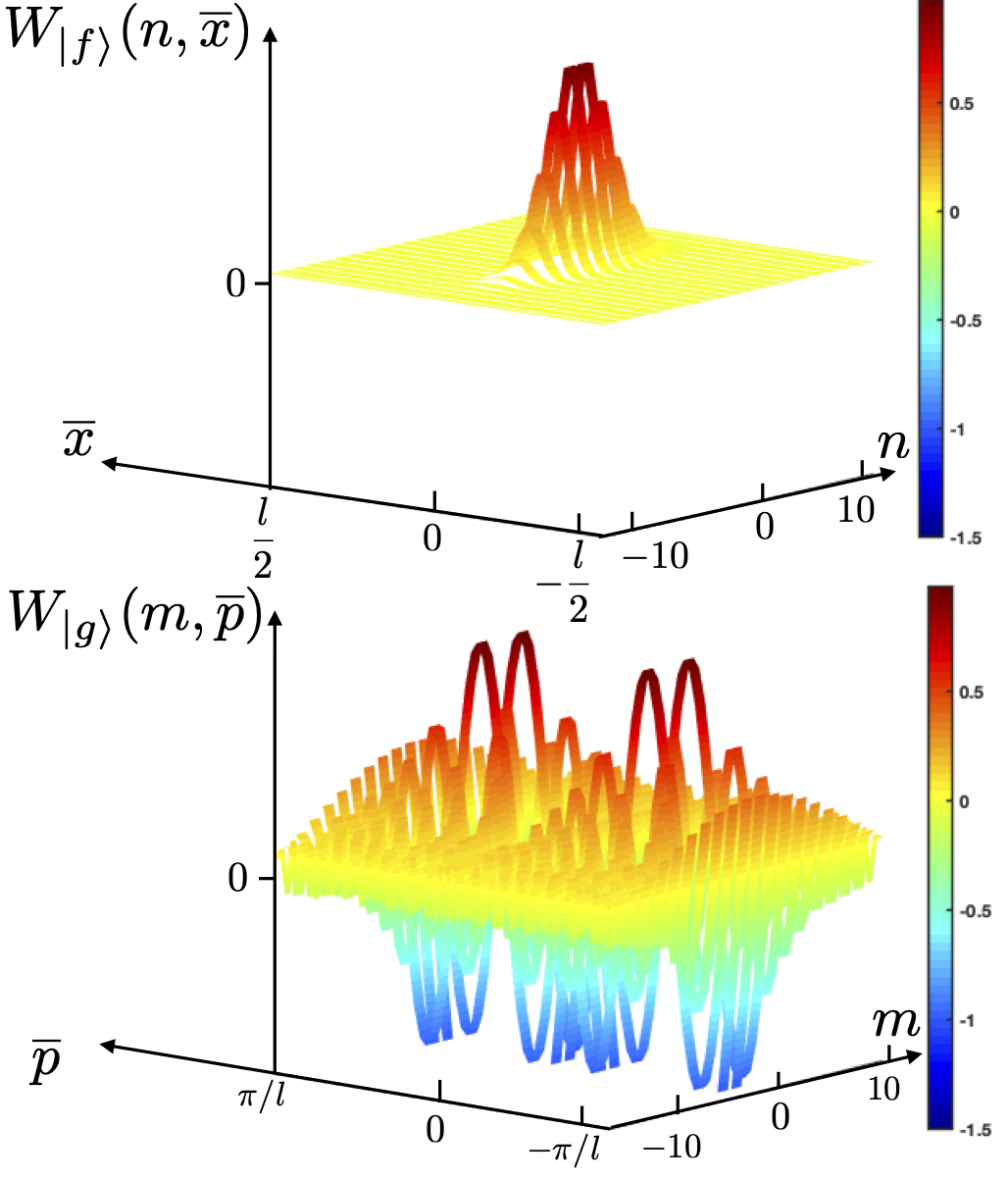}
 \caption{\label{WignermodularGaussian}Modular Wigner distribution of a coherent state in the two cylinders phase space.}
 \end{center}
\end{figure}

The wave function of a coherent state centered in the rectangular phase space at position $x_0$ and momentum $p_0$ can be written under the form:
\begin{equation}
\ket{\psi}=\int_{\mathds{R}} \frac{\text{d}x}{\sqrt{2\pi\sigma^{2}}} e^{-\frac{(x-x_{0})^{2}}{2\sigma^{2}}}e^{ip_{0}(x-x_{0})} \ket{x} 
\end{equation}
where $\sigma$ is the width of the Gaussian function. In the modular basis, the corresponding modular wave function is \cite{Andreasthesis}:
\begin{equation}\label{wrappedgaussian}
\phi(\overline{x},\overline{p})=G_{\sigma}(\overline{x})\Theta_{3}(\frac{l(\overline{p}-p_{0})}{2}-i\frac{l(\overline{x}-x_{0})}{2\sigma^{2}}, G_{\sigma}(l))
\end{equation}
where $l$ the length of the lattice and $\Theta_{3}$ is the third elliptic theta-function: 
\begin{equation}
\Theta_{3}(x,y)=\sum_{n\in\mathds{Z}} y^{n^{2}} e^{-2inx}
\end{equation}
This distribution is also called the wrapped Gaussian distribution. The state is not separable into two functions in the modular variable $\overline{x},\overline{p}$: the two cylinder phase space $(n,\overline{x})$ and $(m,\overline{p})$ are now coupled.  
We can study the state in two limits, choosing the values $x_{0}=p_{0}=0$. In the case where $l\gg \sigma$, the Gaussian state in the modular basis can be approximated by:
\begin{equation}\label{firstlimit}
\phi(\overline{x},\overline{p})\simeq N(\sigma,l) G_{\sigma}(\overline{x}) H_{2\pi/l}(\overline{p})
\end{equation}
whereas in the other limit $\sigma=l$, 
\begin{equation}
\phi(\overline{x},\overline{p})\simeq  N'(\sigma,l)  G_{\sigma}(\overline{p}) H_{2\pi/l}(\overline{x})
\end{equation}
where $N(\sigma,l) $ and $N'(\sigma,l)$ are two normalization factors and $H_{2\pi/l}(\overline{p})$ is the rectangular function centered at zero of width $2\pi/l$. In this approximate form, the modular wave function is separable $\phi(\overline{x},\overline{p})\simeq f(\overline{x})g(\overline{p})$ and is a localized state in the  modular variable $\overline{x}$ but delocalized in the other $\overline{p}$. It can be understood here as a consequence of the non-periodicity property of the amplitude wave function of the state in the $x$-representation. The associated modular Wigner distribution of the state Eq.~(\ref{firstlimit}) is separable  $W_{\ket{\psi}}(n,m,\overline{x},\overline{p})=W_{\ket{f}}(n,\overline{x})W_{\ket{g}}(m, \overline{p})$ and can be written as:
\begin{equation}\label{gaussianmodular}
W_{\ket{f}}(n,\overline{x})\simeq e^{-\frac{\overline{x}^{2}}{2\sigma^{2}}}e^{-n^{2}(2\pi\sigma/l)^{2}}
\end{equation}
\begin{equation}\label{gaussianmodular2}
W_{\ket{g}}(m,\overline{p})=\frac{\text{sin}(2m(-\abs{\overline{p}}+\pi/l))}{2m}.
\end{equation}
These functions are represented in Fig.~\ref{WignermodularGaussian}. 

We now study the modular Wigner distribution of the sum of two Gaussian  state centered at $ x_0=\pm l/4$ and $p_{0}=0$ which corresponds to a Schrödinger cat in rectangular phase space. It exhibits an oscillatory pattern in one cylinder phase space, as the GKP state $\ket{\tilde{\pm}}_{x}$ (see Eq.~(\ref{wignergkp})): 
\begin{multline}
W_{\ket{f}}(n,\overline{x})\simeq e^{-\frac{(\overline{x}+l/4)^{2}}{2\sigma^{2}}}e^{-n^{2}(2\pi\sigma/l)^{2}}+e^{-\frac{(\overline{x}-l/4)^{2}}{2\sigma^{2}}}e^{-n^{2}(2\pi\sigma/l)^{2}}\\\pm e^{-n^{2}(2\pi\sigma/l)^{2}}e^{-\frac{\overline{x}^{2}}{\sigma^{2}}}\text{cos}(\frac{n\pi}{\hbar})
\end{multline}
The other cylinder distribution $W_{\ket{g}}(m,\overline{p})$  is given by Eq.~(\ref{gaussianmodular2}) and is different from the one associated to $\ket{\tilde{\pm}}_{x}$  (see Eq.~(\ref{envelopgkp})).

This example illustrates the relevance of representing the distribution of a quantum state in the two cylinder phase space.

\section{Study of Quantum error correction in the double cylinder phase space}\label{sectionsix}
In this section, we investigate the quantum error correction of periodic and non-periodic states using ancilla GKP states. Our study is based on the scheme presented by Glancy and Knill in Ref. \cite{Glancy} and is illustrated by the representation of modular Wigner distribution before and after the correction on the double cylinder phase space.

\subsection{General formulation of the protocol}
\begin{figure}[h!]
\begin{center}
 \includegraphics[scale=0.6]{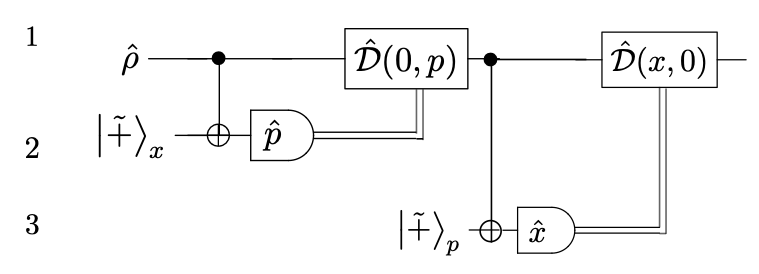}
 \caption{\label{steane}Steane error correction scheme for a general density matrix $\hat{\rho}$ using as ancilla GKP states. The state $\hat{\rho}$ (upper rail) and  the ancilla GKP state $\ket{\tilde{+}}_{x}$ are entangled by a $\hat{C}_{Z}$ gate  followed by a homodyne detection on the spatial port $2$. The measurement result $p$ is used to perform a displacement operator denoted by ${\cal{\hat{D}}}(0,p)$. The procedure is then repeated but using as ancilla state $\ket{\tilde{+}}_{p}$ in spatial port $3$ and a homodyne detection along the $x$-quadrature.  }
 \end{center}
\end{figure}

The quantum error correction of a density matrix $\hat{\rho}$, which can be a noisy GKP or a Gaussian state, using an ancilla GKP state as resource, is as follows. We start by coupling  $\hat{\rho}$ to a GKP ancilla state $\ket{\tilde{+}}_{x}$ with the entangling operation $\hat{C}_{z}=e^{i\hat{x}_{1}\hat{x}_{2}}$, where $\hat{x}_{1/2}$ is the  position operator and $1$ and $2$ denote each spatial port (see Fig.~\ref{steane}). Then a homodyne measurement is performed on the $p$-quadrature on the spatial port $2$.
The state after these steps is:
\begin{equation}\label{aftercorrectiondensity}
\hat{\rho}'=\frac{\bra{p}\hat{C}^{-1}_{z}\hat{\rho}\hat{C}_{z}\ket{p}}{\text{Tr}(\bra{p}\hat{C}_{z}\hat{\rho}\hat{C}_{z}\ket{p})},
\end{equation}
where $p$ is the value measured by the homodyne detection. As specified on Fig.~\ref{steane}, an additional displacement gate ${\cal{\hat{D}}}(0,p)$ can be applied using the measurement result $p$ of the homodyne measurement. In the case of an initial pure state $\hat{\rho}=\ket{\psi}\bra{\psi}$ with $\ket{\psi}=\int \text{d}x \psi(x) \ket{x}$, the amplitude of the wave function in the $x$-representation $\psi_{p}(x)$ after the three steps using Eq.~(\ref{aftercorrectiondensity}) is:
\begin{equation}\label{amplitudeaftercorrection}
\psi_{p}(x)=\psi(x+p)\tilde{\psi}_{\tilde{+}}(x),
\end{equation}
where $\tilde{\psi}_{\tilde{+}}(x)$ is the amplitude of the ancilla state in the $p$-representation given by a Gaussian comb. This result is demonstrated in the Appendix \ref{demonstration}. This protocol  can be seen as the application of shifts errors (the displacement $p$), which update the distribution of the noise described by $\psi(x+p)$ of the initial state and then projecting back the state in the GKP subspace by multiplying $\psi(x+p)$ by $\tilde{\psi}_{\tilde{+}}(x)$. Similar results expressed in terms of Wigner distribution have been obtained in \cite{gkpkraus}.
Correction of the orthogonal quadrature is also possible repeating the protocol but using the ancilla state $\ket{\overline{+}}_{p}=\ket{\overline{0}}_{x}$ in the spatial port $3$ (see Fig.~\ref{steane}) and performing a homodyne measurement in the orthogonal quadrature $\hat{x}$.  In practice, when the initial state is a GKP state, these two steps have to be repeated to fully correct the state with different strategies as the one developed in \cite{Yang} or with a Bayesian optimization procedure \cite{bayesian}.\\

In the following, we suppose for simplicity that the momentum $p$  measured by the homodyne detection is equal to zero and the initial amplitude probability $\psi(x)$ is hence projected into $\psi_{0}(x)=\psi(x+0)\tilde{\psi}_{\tilde{+}}(x)$. Our results can be generalized for different values of $p$  by knowing the probability distribution of $p$.

\subsection{Quantum Error correction of a GKP state using a GKP state as ancilla}\label{gkpcorrectionsection}
\begin{figure*}
\begin{center}
 \includegraphics[scale=0.42]{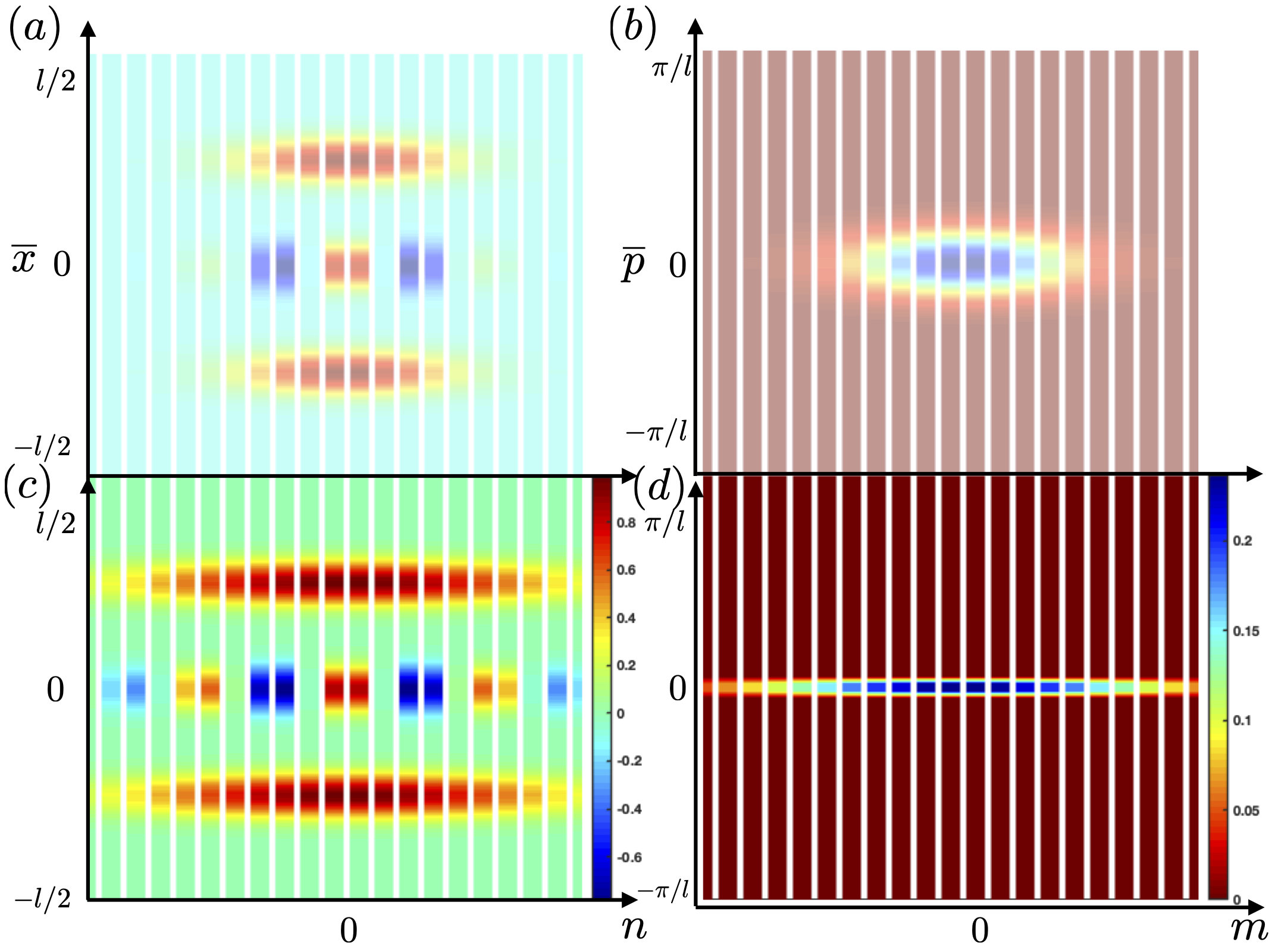}
 \caption{\label{WignerGKPcorr}Illustration of the correction of a noisy data GKP state (a) and (b) representing the modular Wigner distribution. Projection of a GKP state on a slightly less noisy subspace (c) and (d) after the correction procedure. After the second homodyne detection, the probability of having an error less than than $\sqrt{\pi}/6$ goes from 0.9 to 0.99 as the number of oscillations goes from 3 to 5, see Fig.~\ref{visibility}. }
 \end{center}
\end{figure*}

 In Fig.~\ref{WignerGKPcorr}, we present the modular Wigner distribution of the state before the error correction (a), (b) (see also Eq.~(\ref{maingkp}), Eq.~(\ref{envelopgkp})) and after error correction Fig.~\ref{WignerGKPcorr}(c), (d).\\
 
 The two phase spaces of the two separable GKP state  are initially decoupled, and each  GKP states have independent noises in each variable and consequently each phase space. The gate $\hat{C}_{z}$ entangles both qubits and hence couples the two cylinder phase space of  qubits and transfer shift errors from the ancilla qubit to the data one.
Successive homodyne detection will lead to the squeezing of the $\overline{x}$ and $\overline{p}$ and a broadening along the integer direction $n$ and $m$.

\begin{figure}[h!]
\begin{center}
 \includegraphics[scale=0.45]{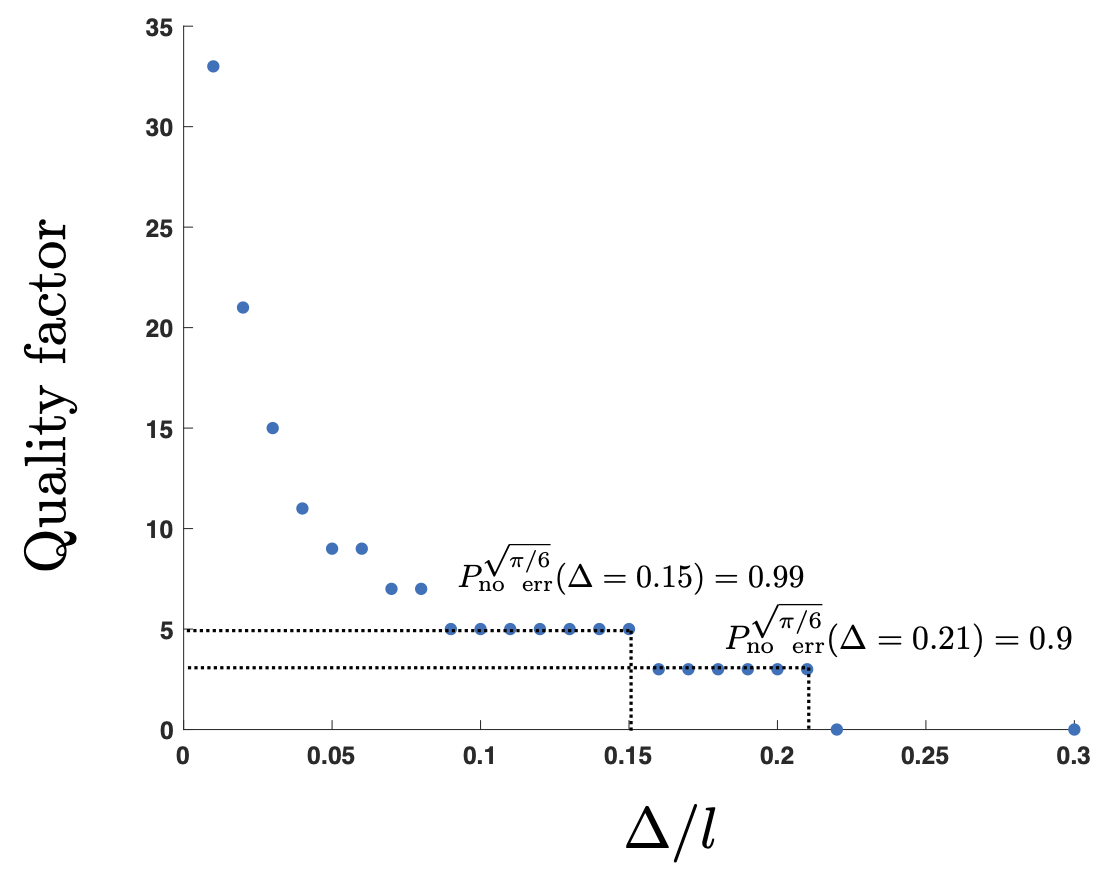}
 \caption{\label{visibility} Quality factor (number of oscillations in the cylinder phase space) with respect to the ratio $\Delta/l$ for $l=\sqrt{\pi}$. After one homodyne detection along the $p$-quadrature for a GKP state, the number of oscillations of the GKP states increase (and so as the quality factor). If the GKP ancilla state is an ideal one, the number of oscillations becomes infinite.}
 \end{center}
\end{figure}

An important figure of merit is the number of oscillations with respect to the ratio $\Delta/l$. As the ratio $l/\Delta\rightarrow 0$ the two GKP logical states overlap and the oscillations disappears. On the contrary,  the number of oscillations goes to infinity when $\Delta/l \rightarrow 0$, {\it{i.e}} when we consider ideal GKP states which are translationally invariant (see the previous section and Eq.~(\ref{wignergkp})).   The number of oscillations is evaluated numerically as a function of the ratio $\Delta/l$ and is represented on Fig.~\ref{visibility}.

In \cite{Glancy}, the authors have developed a figure of merit which indicates the error tolerance of GKP states which have undergone a distribution of shift errors of amplitude $\abs{u}=\abs{\overline{x}},\abs{v}=\abs{\overline{p}}<\sqrt{\pi}/6$. Assuming an initial separable modular wave function, with  $\Delta<0.4$, the probability to have an error less than $\sqrt{\pi}/6$ is given by:
\begin{equation}
P_{\text{no \ err}}^{\sqrt{\pi/6}}(\Delta)=\int^{\sqrt{\pi/6}}_{-\sqrt{\pi}/6} \text{d}\overline{p} \int^{\sqrt{\pi/6}}_{-\sqrt{\pi}/6} \text{d}\overline{x} \abs{\psi(\overline{x},\overline{p})}^{2}
\end{equation}
with $\abs{\psi(\overline{x},\overline{p})}^{2}=G_{\Delta}(\overline{x})G_{\kappa}(\overline{p})$. 
Numerically, for a mean number of photon $\overline{n}=22.1$, and $\overline{n}=10.4$  it corresponds to $P_{\text{no \ err}}^{\sqrt{\pi/6}}(\Delta=0.15)=0.99$ and $P_{\text{no \ err}}^{\sqrt{\pi/6}}(\Delta=0.21)=0.9$ respectively.  

We report these values in Fig.~\ref{visibility}. The correction of physical GKP states permits to approach the ideal GKP states which have an infinite number of oscillations in one cylinder phase space $(\overline{x},n)$.

\subsection{Correction of non-periodic state using GKP state as ancilla}
\begin{figure*}
\begin{center}
 \includegraphics[scale=0.4]{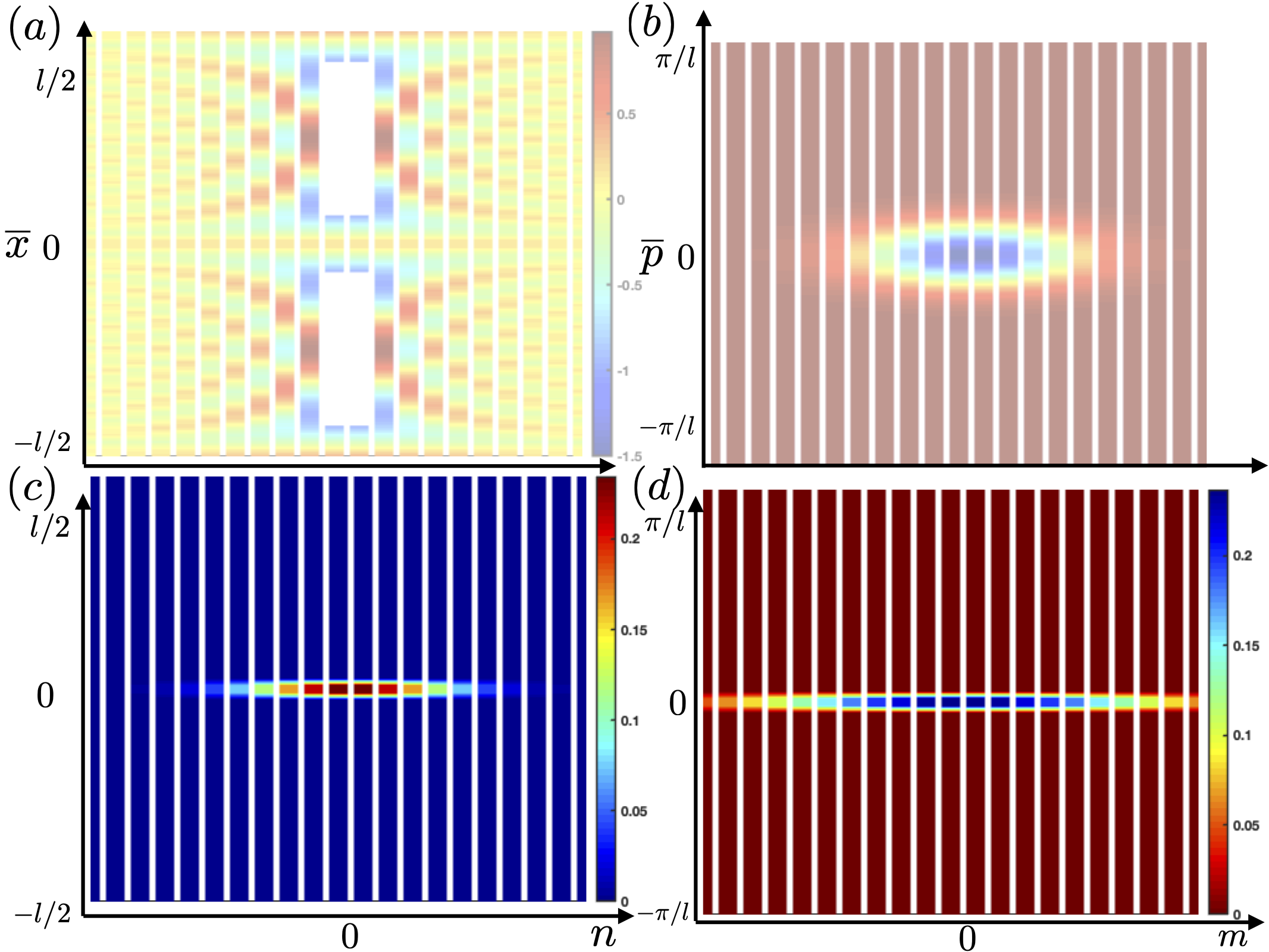}
 \caption{\label{Wignergausscorr}Illustration of the projection of a coherent state (a), (b) on a corrected GKP subspace (c), (d) in the double cylinder phase space.}
 \end{center}
\end{figure*}

We now briefly discuss the case of a coherent state described by Eq.~(\ref{gaussianmodular}) and represented in Fig.~\ref{Wignergausscorr}. (a),(b). The Steane error correction using as an ancilla the state $\ket{\tilde{+}}_{x}$ results to the projection of the coherent state $\hat{\rho}$ on the GKP subspace. After the protocol (see Fig.~\ref{Wignergausscorr}.(c),(d)), the initial coherent state  becomes a random state on this subspace and can be used as a magic state to elevate GKP Clifford QC to fault-tolerant universal QC \cite{gkpkraus}. The protocol is a magic state distillation \cite{magicstate,distillation}: the gate implemented corresponds to a quantum operation outside the set of  Gaussian operations. The state produced is a non-Gaussian one and by consequence is hard to simulate by classical means. The non-Gaussian resource comes from the ancilla GKP state itself and acts as a position-momentum filter.

\section{Proposal of tomographical reconstruction of the Modular Wigner distribution}\label{sectionseven}

\begin{figure}[h!]
\begin{center}
 \includegraphics[scale=0.55]{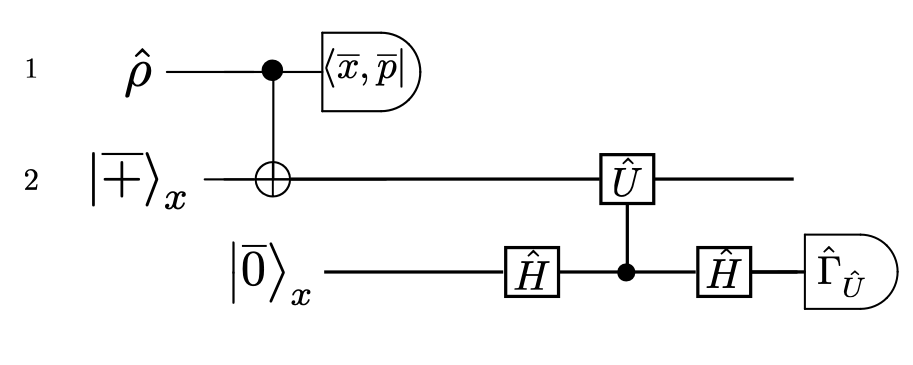}
 \caption{\label{tomographymodular} Tomography of the modular Wigner distribution. The first step is to entangled the state with a GKP ancilla and perform a post-selection on the modular state $\bra{\overline{x},\overline{p}}$. An indirect measurement is performed using an ancilla  GKP qubit $\ket{\overline{0}}_{x}$ to measure the readout modular observable $\hat{\Gamma}_{\hat{U}}$ where $\hat{U}=\hat{X},\hat{Y}$.}
 \end{center}
\end{figure}

In this section, we propose a theoretical scheme to measure the modular Wigner distribution using results developed in \cite{wigneroam} for measuring Wigner distribution for the azimuthal structure of light. The protocol starts with a separable density matrix $\hat{\mu}=\hat{\rho}\otimes\ket{\overline{+}}\bra{\overline{+}}$ where $\hat{\rho}$ is the state to measure and   $\ket{\overline{+}}\bra{\overline{+}}$ is an ancilla ideal GKP state that here will be used as a pointer. The two states are then entangled by the gate $\hat{C}(\alpha,\beta)$, defined as:
\begin{equation}\label{entangledgate}
\hat{C}(\alpha,\beta)=e^{i(\alpha\hat{N}+\beta\hat{M})\otimes\hat{\Gamma}_{z}},
\end{equation}
where $\alpha \in \mathds{S}^{1}$ and $\beta\in\mathds{S}^{1*}$. The operators $\hat{N}$ and $\hat{M}$ are defined by Eq.~(\ref{integeroperator}) and is applied on the spatial port $1$ (see Fig.~\ref{tomographymodular}). $\hat{\Gamma}_{z}$ is the modular readout observable defined by Eq.~(\ref{sigmaz}) and Eq.~(\ref{modulareadout}) applied on the spatial port $2$. After the entangling gate, if we take as an example a pure state $\hat{\rho}=\ket{\psi}\bra{\psi}$, the quantum state $\hat{\mu}_{\alpha,\beta}=\hat{C}(\alpha,\beta)\hat{\mu}\hat{C}^{-1}(\alpha,\beta)=\ket{\mu}_{\alpha,\beta}\bra{\mu}_{\alpha,\beta}$ becomes a linear superposition,
\begin{equation}
\ket{\mu}_{\alpha,\beta}=\frac{1}{\sqrt{2}}(e^{i(\alpha\hat{N}+\beta\hat{M})}\ket{\psi} \otimes \ket{\overline{0}}+e^{-i(\alpha\hat{N}+\beta\hat{M})}\ket{\psi} \otimes\ket{\overline{1}}).
\end{equation}
A post-selection measurement on the modular state $\bra{\overline{x},\overline{p}}$ gives the following reduced density matrix:
\begin{equation}\label{statepostselect}
\hat{\mu}'_{\alpha,\beta}=\frac{\bra{\overline{x},\overline{p}}\hat{\mu}_{\alpha,\beta}\ket{\overline{x},\overline{p}}}{\text{Tr}(\bra{\overline{x},\overline{p}}\hat{\mu}_{\alpha,\beta}\ket{\overline{x},\overline{p}}}.
\end{equation}
In the case of the pure case considered here, the numerator of the previous equation can be written as: $\frac{1}{\sqrt{2}}(\bra{\overline{x}+\alpha,\beta+\overline{p}}\ket{\psi} \otimes \ket{\overline{0}}+\bra{\overline{x}-\alpha,\overline{p}-\beta}\ket{\psi} \otimes\ket{\overline{1}})$. We point out that the real and imaginary part of the correlation function
\begin{equation}
C(\overline{x},\overline{p},\alpha,\beta)=\bra{\overline{x}+\alpha,\overline{p}+\beta}\hat{\rho}\ket{\overline{x}-\alpha,\overline{p}-\beta},
\end{equation}
 can be obtained by measuring the readout modular observables ($\hat{\Gamma}_{x}$ and $\hat{\Gamma}_{y}$ defined in  Eq.~(\ref{modulareadout})) of the ancilla qubit. Indeed, thanks to Eq.~(\ref{statepostselect}), the expectation value of the modular readout observables are
\begin{align}
\langle \hat{\Gamma}_{x} \rangle=  \text{Re}(\bra{\overline{x}+\alpha,\overline{p}+\beta}\hat{\rho}\ket{\overline{x}-\alpha,\overline{p}-\beta}),\\
\langle \hat{\Gamma}_{y} \rangle=  \text{Im}(\bra{\overline{x}+\alpha,\overline{p}+\beta}\hat{\rho}\ket{\overline{x}-\alpha,\overline{p}-\beta}).
\end{align}
Once $ \hat{\Gamma}_{x},  \hat{\Gamma}_{y}$  have been measured  for the ancilla qubit we can reconstruct  the correlation function of the quantum state of interest. A  final Fourier transform of the expectation value of the modular readout observables permits to obtain the Modular Wigner distribution. The measurement of the $\hat{\Gamma}_{\hat{U}}$ matrices is fully detailed in \cite{Andreasthesis}. In Fig.~\ref{tomographymodular}, we present the full quantum circuit that allows to measure the modular Wigner distribution. 
The measurement of the expectation value of the readout modular observables using an indirect measurement strategy is inspired from \cite{Asadian1,Asadian2} and could be performed in various experimental platform, such as spatial degrees of freedom of single photons \cite{spatialcv}, mechanical oscillators and vibrational modes of ions.\\ 

We detail now how to perform the post-selection measurement on the modular basis $\bra{\overline{x},\overline{p}}$, taking as an example a general pure state $\ket{\psi}=\int \text{d} q \psi(q) \ket{q}$. We start by applying two small displacements shifts $\hat{D}(\overline{x})\hat{D}(\overline{p})$ which transform the state as: $\ket{\psi}\rightarrow \int \psi(q+\overline{x}) \ket{q}e^{-i\overline{p}q}\text{d}q$. A projective measurement  which selects the position $q=kl (k\in\mathds{Z})$, that is equivalent to a projection onto  state $\sum_k\bra{q=kl}$, gives the probability: $\abs{\sum_{k} \psi(kl+\overline{x}) e^{-i\overline{p}kl}}^{2}=P(\overline{x},\overline{p})$, which is the absolute value of the modular wave function $\psi(\overline{x},\overline{p})$ (see Eq.~(\ref{modularwave})). It could be experimentally implemented in the context of the transversal degrees of freedom of single photons \cite{SLM}.
An experimental proposal to implement the entangling gate Eq.~(\ref{entangledgate}) will be the subject of future work.

To conclude, the mentioned protocol could be alternatively performed using the controlled displaced operator $\hat{E}'(a,b)=\text{exp}(i(a\hat{\overline{x}}+b\hat{\overline{p}})\otimes\hat{\Gamma}_{Z})$, followed by a post-selection on the basis $\bra{l,m}$ and then a measurement of the read-out modular variable. The choice of the most suitable protocols depends on the type of experimental devices available.

\section{Conclusion}
We introduced a new Wigner distribution well adapted for translational invariant states, based on the tools developed in \cite{Zak,Andreasthesis}. This Modular Wigner distribution is represented in a double cylinder phase space and each cylinder could be coupled or not depending on the considered quantum state and its associated noise model. Superposition of localized states in the modular plane have a Schrödinger cat shape in one cylinder phase space. We have seen that one figure of merit which quantifies the possibility to correct the GKP states is related to the number of oscillation in one cylinder phase space. We hope that this work gives a new framework to implement other quantum information protocols involving discrete symmetry, as the period finding problem \cite{shorcontinuous}.

\begin{appendix}

\section{Commutators }\label{appendixcommu}
In this appendix, we introduce the commutators of the integers operators and the modular ones based on the calculations from Ref. \cite{slitsmodular}. They are not zero in the case where the periodicity of the two lattices defined along the position and momentum axis are $l$ and $2\pi/l$:\begin{equation}
[\hat{\overline{x}},\hat{N}_{p}]=\frac{il}{2\pi}(\mathds{I}-l\int^{\pi/l}_{-\pi/l} \text{d}\overline{p} \ket{l/2,\overline{p}} \bra{l/2,\overline{p}}),
\end{equation}
\begin{equation}
[\hat{N}_{x},\hat{\overline{p}}]= \frac{i}{l}(\mathds{I}-\frac{2\pi}{l} \int_{-l/2}^{l/2} \text{d}\overline{x} \ket{\overline{x},\pi/l}\bra{\overline{x},\pi/l}),
\end{equation}
where the second term on the right-hand side of the two previous equations are a sums of projectors on modular eigenstates. These relations constitute an additional proof of the distinction between the integers operators $\hat{N}_{x}$ and $\hat{N}_{p}$ and the angular momentum operators $\hat{N}$ and $\hat{M}$ defined by Eq.~(\ref{integeroperator}).

\section{Composition of two modular displacement operators}
In this section, we prove  Eq.~(\ref{composition}) of the main text. The product of two displacement operators acting on the integer state $\ket{n_{1},m_{1}}$ gives:
{\small{\begin{multline}
\hat{D}(n,\overline{x},m,\overline{p})\hat{D}(n',\overline{x'},m',\overline{p'})\ket{n_{1},m_{1}}=e^{i\overline{x}(n_{1}+n'+n/2)}\\\cross e^{-i\overline{p}(m_{1}+m'+m/2)}e^{i\overline{x'}(n_{1}+n'/2)}e^{-i\overline{p'}(m_{1}+m'/2)}\\ \cross \ket{n_{1}+n+n',m_{1}+m+m'}
\end{multline}}}
whereas the composition of displacement operators gives:
{\small{\begin{multline}
\hat{D}(n+n',\overline{\overline{x}+\overline{x'}},m+m',\overline{\overline{p}+\overline{p'}}\ket{n_{1},m_{1}}=e^{i\overline{\overline{x}+\overline{x'}}(n_{1}+(n+n')/2)}\\ \cross e^{-i\overline{\overline{p}+\overline{p'}}(m_{1}+(m+m')/2)}\ket{n_{1}+n+n',m_{1}+m+m'}.
\end{multline}}}
Using the relation $\overline{\overline{x}+\overline{x'}}=\overline{x}+\overline{x'} \mp l H(\abs{\overline{x}-l/2\pm \overline{x'}})$ for $\overline{x}\gtrless0$ with  $H$ being the Heaviside function, we recover Eq.~(\ref{composition}) , since $\text{exp}(-2\pi n_{1} H(\overline{x}))=1$ for any $n_{1}$ and $\overline{x}$.

\section{Error correction}
In this appendix, we summarize the derivation of the analytical probability that a GKP state has shift errors smaller than a certain threshold using \cite{Glancy}. We also develop the calculation of the amplitude coefficient  GKP state in the integer representation as well as the amplitude wave function in the position representation of the state after the Steane error correction procedure.

\subsection{Shift errors}
Here we recall the calculation of the probability that physical GKP state has shifts smaller than $\sqrt{\pi/6}$ in both quadrature \cite{Glancy}. If the ideal $\ket{\overline{0}}$ logical state has errors shifts $u$, $v$ in position and momentum quadrature, it can be written as:  
\begin{equation}
\ket{u,v}=e^{-iv\hat{q}}e^{iu\hat{p}}\ket{\tilde{0}}.
\end{equation}
This basis actually corresponds to the modular basis described by Eq.~(\ref{Zakt}), only when $\abs{u}<\sqrt{\pi}$ and $\abs{v}<\sqrt{\pi}$. From this consideration, we can develop any wave function into that basis:
\begin{equation}
\ket{\psi}=\int^{\sqrt{\pi}}_{-\sqrt{\pi}}\int_{-\sqrt{\pi}/2}^{\sqrt{\pi}/2} \text{d}v\text{d}u \psi(u,v) \ket{u,v},
\end{equation}
where $\psi(u,v)$ corresponds to the modular wave function Eq.~(\ref{modularwave}). The probability of having an error $u,v$ is $P(u,v)=\abs{\bra{\psi}\ket{u,v}}^{2}$ and the probability of having an error less than $\sqrt{\pi/6}$ is given by:
\begin{equation}
P_{\text{no \ err}}^{\sqrt{\pi/6}}(\Delta)=\int^{\sqrt{\pi/6}}_{-\sqrt{\pi}/6} \text{d}\overline{p} \int^{\sqrt{\pi/6}}_{-\sqrt{\pi}/6} \text{d}\overline{x} \abs{\psi(\overline{x},\overline{p})}^{2},
\end{equation}
and is evaluated numerically for different average value of the photon number $\overline{n}\sim \frac{1}{2\Delta^{2}}$ in Sec. \ref{gkpcorrectionsection}, $\Delta$ being the variance of the Gaussian distribution of both variables $u$ and $v$.

\subsection{Physical GKP states in the integer basis}\label{GKPinteger}
In this section, we calculate the physical GKP state $\ket{\tilde{0}}$ in the integer basis $\bra{n,m}\ket{\tilde{0}}=f_{n}g_{m}$, where
\begin{equation}\label{FN}
f_{n}=\int_{-l/2}^{l/2} \text{d}\overline{x} e^{-\frac{2i\pi}{l}n\overline{x}} e^{-(\overline{x}-l/4)^{2}/(2\Delta^{2})}
\end{equation}
and 
\begin{equation}\label{GN}
g_{m}=\int_{-\pi/l}^{\pi/l} \text{d}\overline{p} e^{-iml\overline{p}} e^{-\overline{p}^{2}/2\kappa^{2}}.
\end{equation}
We first calculate Eq.~(\ref{FN}). After performing a change of variable, we obtain:
\begin{align}
f_{n}=e^{i\pi n/2} \int_{-l/4}^{3l/4} \text{d}\overline{x} e^{-\frac{2i\pi}{l}n\overline{x}} e^{-\overline{x}^{2}/2\Delta^{2}}\\
= e^{i\pi n/2} e^{-2(\frac{\pi n\Delta}{l})^{2}} \int_{-l/4}^{3l/4} \text{d}\overline{x} e^{-(\overline{x}+2i\pi n\Delta^{2}/l)/2\Delta^{2}}
\end{align}
We hence recognize the error function $\text{erf}(l/2\Delta)=\int_{-l/2}^{l/2}e^{-\overline{x}^{2}/2\Delta^{2}} \text{d}\overline{x}$,  with a complex argument. The integer coefficient of the physical GKP state is
\begin{multline}
f_{n}=\sqrt{\frac{\pi}{2}}e^{i\pi n/2} e^{-2(\frac{\pi n\Delta}{l})^{2}}[\text{erf}(\frac{l}{4\sqrt{2}\Delta}-\frac{\sqrt{2}in\pi\Delta}{l})\\+\text{erf}(\frac{3l}{4\sqrt{2}\Delta}+\frac{\sqrt{2}in\pi\Delta}{l})].
\end{multline}
A similar calculation leads to the other coefficient Eq.~(\ref{GN}):
\begin{equation}
g_{m}=e^{-(ml\kappa)^{2}/2}[\text{erf}(\frac{\pi}{\sqrt{2}\kappa l}-im\kappa l/\sqrt{2})+\text{erf}(\frac{\pi}{\sqrt{2}\kappa l}+im\kappa l/\sqrt{2})].
\end{equation}
The $\ket{\tilde{1}}$ physical GKP state can be expressed in the integer basis using similar calculation.

\subsection{Wave function of the state Eq.~(\ref{amplitudeaftercorrection}) after the error correction protocol}\label{demonstration}
In this section, we demonstrate Eq.~(\ref{amplitudeaftercorrection}). We start from an initial pure state $\hat{\rho}=\ket{\psi}\bra{\psi}$, with wave function $\ket{\psi}=\int \text{d}x \psi(x)\ket{x}$. The state  is entangled  with the ancilla physical GKP state $\ket{\tilde{+}}=\int \psi_{\tilde{+}}(x_{2}) \ket{x_{2}} \text{d}x_{2}$ with the entangling gate $\hat{C}_{z}=e^{i\hat{q}_{1}\hat{q}_{2}}$.
In the position representation, the total wave function can be written as:
\begin{equation}
\ket{\psi}=\iint e^{ix_{1}x_{2}}\psi(x_{1})\psi_{\tilde{+}}(x_{2}) \ket{x_{1}}\ket{x_{2}} \text{d}x_{1}\text{d}x_{2}.
\end{equation}
An homodyne detection along the $p$-quadrature is performed on the ancilla port, the resulting state after such detection becomes:
\begin{equation}
\ket{\psi}=\iint e^{ix_{1}x_{2}}e^{ix_{2}p}\psi(x_{1})\psi_{\tilde{+}}(x_{2}) \ket{x_{1}}\text{d}x_{1}\text{d}x_{2}.
\end{equation}
After integration over the variable $x_{2}$, we obtain:
\begin{equation}
\ket{\psi}=\int \psi(x_{1})\tilde{\psi}_{\tilde{+}}(x_{1}+p)\ket{x_{1}} \text{d}x_{1},
\end{equation}
where $\tilde{\psi}_{\tilde{+}}(p)=\int e^{ixp} \psi_{\tilde{+}}(x) \text{d}x$. It proves Eq.~(\ref{amplitudeaftercorrection}).

\end{appendix}

\bibliography{modularwigner2}

\end{document}